\documentclass[aps,prb,groupedaddress,twocolumn,showpacs]{revtex4}

\usepackage{relsize}
\usepackage{amsmath,amssymb,bbm}
\usepackage{graphicx}

\def\myTens#1{ {\vbox{\hbox{$\smash {#1}$} \vskip0.3ex \hrule \vskip0.2ex \hrule \vskip-0.68ex }} }
\def\myTensS#1{ {\vbox{\hbox{$\smash {#1}$} \vskip0.2ex \hrule \vskip-0.3ex }} }

\newcommand{\Tr}{\mathrm{Tr}}
\newcommand{\upd}{ {\rm d} }
\newcommand{\abs}[1]{{\left|#1\right|}}
\newcommand{\avg}[1]{\left\langle#1\right\rangle}
\newcommand{\mv}[1]{\left\langle#1\right\rangle}

\newcommand{\vect}[1] { {\boldsymbol{#1}} }

\newcommand {\dyadic} { \circ }

\newcommand{\sig}{\hat {\vect \sigma}}

\newcommand{\m}{\vect m}
\newcommand{\rt}{\vect r, t}

\newcommand{\lm}  { l_{\m} }

\newcommand{\xii} {\chi^{(i)}}     
\newcommand{\xim} {\chi^{(m)}}     
\newcommand{\ximTens} {\myTens{\chi}^{(m)}}
\newcommand{\xims} {\xim_\perp}     
\newcommand{\ximp} {\xim_\parallel} 

\newcommand{\eF}  { E_{\rm F} }

\newcommand{\pF}  { p_{\rm F} }
\newcommand{\vF}  { v_{\rm F} }

\newcommand{\delmu}  { {\delta\mu} }

\newcommand{\UP} { \uparrow }
\newcommand{\DOWN} { \downarrow }
\newcommand{\UPDOWN} { {\uparrow\!,\downarrow} }

\newcommand{\PU}  { \hat P_\UP }
\newcommand{\PD}  { \hat P_\DOWN }

\newcommand{\de} { \partial_\epsilon }
\newcommand{\dr} { \partial_{\vect r} }
\newcommand{\dk} { \partial_{\vect k} }

\newcommand{\Gless}  {\hat G^<}
\newcommand{\Sless}  {\hat \Sigma^<}

\newcommand{\ls} { l_{\rm s} }         
\newcommand{\lsd} { l_{\rm sd} }       
\newcommand{\lprec} { l_{\rm prec} }   

\newcommand{\calA}{\mathcal{A}}

\newcommand{\calF}{\mathcal{F}}

\newcommand{\calH}{\mathcal{H}}
\newcommand{\calI}{\mathcal{I}}
\newcommand{\calJ}{\mathcal{J}}

\newcommand{\calP}{\mathcal{P}}

\newcommand{\calS}{\mathcal{S}}
\newcommand{\calT}{\mathcal{T}}
\newcommand{\calV}{\mathcal{V}}

\begin{document}

\title{Electronic Transport in Ferromagnetic Conductors with
  Inhomogeneous Magnetic Order Parameter - Domain-Wall Resistance} 
\author{Christian Wickles}
\email[]{christian.wickles@uni-konstanz.de}
\affiliation{Universit\"at Konstanz, Fachbereich Physik, 78457
  Konstanz, Germany}
\author{Wolfgang Belzig}
\email[]{wolfgang.belzig@uni-konstanz.de}
\affiliation{Universit\"at Konstanz, Fachbereich Physik, 78457
  Konstanz, Germany}

\date{\today}

\begin{abstract}
  We microscopically derive transport equations for the conduction
  electrons in ferromagnetic materials with an inhomogeneous
  magnetization profile. Our quantum kinetic approach includes elastic
  scattering and anisotropic spin-flip scattering at magnetic
  impurities. In the diffusive limit, we calculate the resistance
  through a domain wall and find that the domain-wall resistance can
  be positive or negative. In the limit of long domain walls we derive
  analytical expressions and compare them with existing works, which
  used less general models or different theoretical frameworks.
\end{abstract}



\maketitle

\section{\label{sec:introduction}Introduction}
Conducting magnetic materials are an active research topic at present
due to promising applications like magnetic memory storage devices
which make use of magnetization reversal in pillar multilayer nanostructures
\cite{Myers1999,Urazhdin2003,Ozyilmaz2003,Grollier2001,Katine2000}
or domain wall motion\cite{Tatara2004,Li2004,Zhang2004,Thiaville2005,Barnes2005,Barnes2006}
as proposed for the racetrack memory \cite{Parkin08}.
On one hand, domain wall motion is realized by
sending spin-polarized current through the domain wall, so that the
mutual interaction of the electron spin with the ferromagnetic order
parameter leads to  a motion of the wall. This is due to the so called
spin-torque \cite{Slon96,Ber96,Tsoi98,Myers99,Tser05}, the transfer of spin-angular momentum.
On the other hand,
the electronic current flow is also affected by the presence of an
inhomogeneous magnetization. Most
prominently, there is a change in the resistance when the current runs
through a domain wall in comparison to the resistance in the absence
of the domain wall. The resistance change can have different origins
that can be seperated into the {\it extrinsic} and {\it intrinsic}
domain-wall resistance (DWR). The former includes
orbital and anisotropic magneto-resistance. The latter
contains the direct influence the domain wall has on the electronic
conduction channels: if the magnetization direction is not homogeneous
in space, the spin majority and minority channels are no longer
eigenstates, which in turn changes the conduction properties and also
can have influence on the impurity scattering rates. There is also
spin accumulation in the vicinity of the domain wall which leads to an
additional potential drop.  In any case, the {\it extrinsic}
mechanisms have to be carefully identified in order to obtain the
intrisic domain-wall resistance from experiment. The DWR has been
studied in a number of works in the past, both theoretically \cite{Levy1997, Tatara1997,
  Brataas1999, Gorkom1999, Tatara2001, Dugaev2002, Bergeret2002,
  Simanek2001, Simanek2005} and experimentally
\cite{Ruediger1998,Ebels2000,Aziz2006,Hassel2006}.
Reviews about DWR in nanowires made from ferromagnetic transition
metals, experimental measurements and details on the treatment of
extrinsic magneto resistance can be found in
\cite{Marrows2005,Kent2001}.

On the theoretical side, several limiting case have been investigated
using a variety of theoretical methods.  The works
\cite{Tatara1997,Brataas1999,Gorkom1999} perform a diagrammatic
evaluation of the Kubo-formula introducing scattering in the
unperturbed Greens functions by two phenomenological parameters
$\tau_{\uparrow\!\downarrow}$, the momentum scattering times for
spin up and down channels. In this calculation, spin-flip
processes are not included. As we will discuss later, this leads to a
spin accumulation that does not decay even arbitrarily far from the
domain wall. Hence, this neglect of spin-flip is only possible, if the
distance between the domain wall and leads is much smaller than the
spin-diffusion length. In a complementary approach, Levy and Zhang
\cite{Levy1997} use a linearized Boltzmann equation. They do not
consider changes in the electronic spectrum, i.e. they assume
spin-independence of the wave vector $k_\UP = k_\DOWN$,
restricting the validity to the regime of small exchange
splitting. Their analytical calculation is done in a basis that
diagonalizes the Hamiltonian, which is possible in case of a constant
magnetization gradient, known as spin-spiral. Thus, they cannot take
into account a finite contact geometry and finite domain wall length,
but have to consider an infinitely extended spin-spiral for which they
calculate the conductivity. Spin-flip processes are absent, so, again,
the above statement concerning spin accumulation applies. Furthermore,
they perform a multi-pole expansion of the distribution function but
only include terms up to the p-wave component. However, as we will see
during our calculation, this is not sufficient in general. Lastly, we
believe the Boltzmann equation, they use, lacks terms that should
appear as a result of the gauge transformation.  Bergeret {\it et al}
use the Keldysh technique to derive a quasiclassical equation valid in
the diffusive limit \cite{Bergeret2002}. However, they consider a
different regime of validity, in which the scattering mean free path $\ls$
is the smallest length scale in the system (besides the Fermi-length),
and not the precession length as will be the case in our treatment. Likewise, they
do not consider spin-flip processes, even though during their
calculation, they perform steps which implicitly require longitudinal
spin excitations to relax. Finally, Simanek {\it et al.}
\cite{Simanek2001,Simanek2005} used equations of motion for the
quantum distribution function in Wigner space, which however contain a
term that we cannot reproduce. Before, Bergeret {\it et al.} noted
that this term violates particle conservation
\cite{Bergeret2002}. Nevertheless, this term does not affect the
statement of Simanek {\it et al.} that there is quenching of the
spin-accumulation due to rapid transverse precession. This also emerges
from our theory and we will make use of it later (see the discussion around 
Eq. (\ref{eq:transverseSolution})).

In this article, we pursue a fully microscopical theoretical approach
to the DWR in the limit of wide walls, so that quantum mechanical
electron reflection at the domain wall can be neglected. This allows
us to use a standard quasiclassical approximation and neglect
spin-dependent scattering due to abrupt potential changes
\cite{Huertas2002,Cottet2005,Huertas2000}. In section II, we begin by introducing
our model and deriving a quantum transport equation using the Keldysh
kinetic equation approach. These provide a rather general framework to
investigate a large variety of static transport problems. In Section
III we solve these resulting equations analytically in certain
limiting cases for model domain walls and discuss our results and
relate them to various existing theoretical works dealing with the
issue of DWR \cite{Levy1997, Tatara1997, Brataas1999, Gorkom1999,
  Tatara2001, Dugaev2002, Bergeret2002, Simanek2001,
  Simanek2005}. Finally we conclude with an outlook on open problems.

\section{Quantum Transport Equation for Ferromagnetic Conductors}
In this section we derive a quantum transport equation from a model
Hamiltonian that describes the kinetics of conduction electrons in
materials with inhomogeneous magnetization profile.

\subsection{Model and Hamiltonian}
We consider a system of effectively non-interacting electrons whose
spin degrees of freedom are coupled to the ferromagnetic order
parameter in the mean-field approximation via the spatially dependent
exchange field.  The single particle Hamiltonian has three
contributions,
\begin{eqnarray}
  \hat \calH &=& \hat \calH_0 + \hat \calH_{\rm S} + \hat \calV \\ \nonumber
  &=& \left(\frac {\hbar^2 \vect k^2} {2M} - \delmu(\vect r) -
    e\varphi(\vect r)\right) \hat 1 - \frac\Delta2 \m(\vect r) \sig +
  \hat \calV(\vect r) 
\end{eqnarray}
Here, $\calH_0$ is the usual free quasi-particle energy contribution
with the dispersion relation $\epsilon_{\vect k} \equiv \frac {\hbar^2
  \vect k^2} {2M}$ and effective mass $M$ and in spatial
representation, $\vect k = -i\vect \nabla_{\vect r}$. $\varphi(\vect
r)$ is the external electric potential felt by the quasi-particles of
charge $-e$. $\delmu$ is a shift in the chemical potential
due to the magnetization gradient $\dr\m$ which later turns out to be
of order $\delmu=O(\dr\m)^2$. $\calH_{\rm S}$ describes the coupling
of the electron spin to the exchange field with a constant magnitude
$\Delta$ and the local magnetization direction denoted by the unit
vector $\m(\vect r)$. The $2\times2$ matrix spin structure is denoted
by $\hat{\ }$. $\sig = (\hat\sigma_1,\hat\sigma_2,\hat\sigma_3)$ is
the vector of Pauli matrices, such that the electron spin operator is
given by $\hat {\vect S} = \frac\hbar2 \sig$.

In accordance with the mean-field approach, the length scale of the
spatial variations is much slower than the relevant atomic
scales. More specifically, this condition reads
\begin{eqnarray}
\label{eq:conditionForMagnetization}
1/\lm \equiv \abs{\partial_r \m} \ll \pF/\hbar \ .
\end{eqnarray}
The exchange field is created by electrons that align their spin
preferably in the same direction due to the (here ferromagnetic)
exchange interaction. In conducting ferromagnets, the electrons
contributing to the local magnetization can either be localized and,
thus, do not participate in transport (d-electron character) or be
delocalized and, hence, are subject to electronic transport phenomena
(dominant s-electron character). These extreme cases constitute two
distinct models with the major difference being the way in which the
self-consistency condition for the exchange field is employed. These
are known as {\it s-d} model and {\it itinerant} Stoner model, the
latter one describing a system in which transport and magnetism arise
in fact both from the same delocalized electrons. However, real
physical systems are usually between these two cases. Below, we will
restrict ourselves to the {\it s-d} model in which the magnetization profile
remains static even if the conduction electrons are in a
non-equilibrium configuration. Note, that fluctuations of the order
parameter are neglected.

We disregard the influence of the effective magnetic exchange field on
the electronic orbits, which represents the Lorentz force and leads to
the orbital magneto-resistance (OMR). Theoretically, as well as
experimentally, the OMR and other extrinsic contributions such as the
AMR (anisotropic magneto-resistance), which stems from spin-orbit
coupling and leads to a resistance depending on the angle between
magnetization and current directions, can be separated from the true
DWR and therefore are not included in this article. For Bloch walls
and the CPW-geometry (current perpendicular to wall), which we employ
in our model later on (see FIG. \ref{fig:contact_with_domain_wall}),
the AMR plays no role, since the magnetization direction is always
perpendicular to the current direction.
For other walls like N\'eel walls, the spin direction
within the wall attains a component parallel to the current and the
AMR has to be carefully distinguished from the DWR.

The impurity scattering potential can be divided into two
contributions,
\begin{equation}
  \label{eq:impurity_potential}
  \hat{\mathcal V}(\vect r) = \hat V_{\rm i}(\vect r) + \hat V_{\rm
    mag}(\vect r) \ . 
\end{equation}
$\hat V_{\rm i}$ describes the scattering from randomly distributed
static impurities and for point-like scatterers has the property
\begin{equation}
  \langle V_{\rm i}(\vect r) V_{\rm i}(\vect r')\rangle_{\rm imp} =
  {\xii} \delta(\vect r - \vect r') \ , 
\end{equation}
equivalent to the treatment of $V_{\rm i}$ as a delta-correlated
fluctuating Gaussian field. $\xii$ measures the strength of the impurity
scattering potential and $\langle\rangle_{\rm imp}$ denotes the
averaging over all impurity configurations.

In a similar way, scattering at impurities that have internal
spin-degrees of freedom makes the scattering vertex spin-dependent,
such that
\begin{equation}
  \langle\hat V_{\rm mag}(\vect r) \dyadic \hat V_{\rm mag}(\vect
  r')\rangle_{\rm imp} = \sum_{i,j=1}^3{\xim_{ij}} \hat \sigma_i \dyadic \hat
  \sigma_j \delta(\vect r - \vect r') \ . 
\end{equation}
Therefore, magnetic impurity scattering in the
point-like limit can be treated as delta-correlated fluctuating
Gaussian magnetic field which couples to the electron spin by the usual Zeeman term.
The size of the fluctuations can be generally spin-anisotropic which manifests
itself in the tensor structure of ${\xim_{ij}}$.

\subsection{Keldysh Technique}
The standard way to proceed in non-equilibrium physics is to set up
the kinetic equation for the Keldysh Greens function. For the
remaining part of this article, we set $\hbar = 1$.

\newcommand{\partialLeft}[1]{ \stackrel{\leftarrow}{\partial}_{#1} }
\newcommand{\partialRight}[1]{ \stackrel{\rightarrow}{\partial}_{#1} }

The further treatment is done in the Wigner-representation which is
obtained from usual spatio-temporal representation via the
transformation
\begin{eqnarray}
  \label{eq:wigner_xform}
  \check G(\vect k,\epsilon,\rt) & = &  \int \upd^3\!z d\tau \  e^{-i\vect k
    \vect z + i\epsilon \tau} \\\nonumber 
  && \times \check G \left (\vect r + \frac {\vect z} 2,t +
  \frac\tau2; \vect r - \frac {\vect z} 2,t - \frac\tau2\right)\,.
\end{eqnarray}
The product of operators has to be carried out using the rule
\begin{widetext}
\begin{equation}
  \label{eq:wigner_product}
  C(\vect k,\epsilon,\rt) =  A(\vect k,\epsilon,\rt) \otimes
  B(\vect k,\epsilon,\rt)
  \equiv A(\vect k,\epsilon,\rt) e^{\frac i 2 ( \partialLeft{\vect r} \partialRight{\vect k}
    - \partialLeft{t} \partialRight{\epsilon} - \partialLeft{\vect
      k} \partialRight{\vect r}
    + \partialLeft{\epsilon} \partialRight{t} )}  B(\vect
  k,\epsilon,\rt) \ , 
\end{equation}
\end{widetext}
where $\partialLeft{}$ and $\partialRight{}$ denote derivatives acting
only to the left and right, respectively. The transformation
(\ref{eq:wigner_xform}) introduces center of mass coordinates $\vect
r$, $t$ and Fourier transformed relative coordinates $\vect k$ and
$\epsilon$, respectively. Note, that the product $\otimes$ is associative.

The Greens function $\check G$ is defined as expectation value of the
electron field operators, time-ordered along the Keldysh contour
\cite{Rammer1986}.  The ordering along backward and forward time
Keldysh contour gives rise to an additional $2\times2$ matrix
structure (denoted by $\check{\ \ }$), which in an appropriate basis
takes the convenient form
\begin{equation}
\check G = \left( \begin{array}{cc} \hat G^{\rm R} & 
\Gless \\ 0 & \hat G^{\rm A} \end{array} \right) \ .
\end{equation}
$\hat G^{\rm R}$ and $\hat G^{\rm A}$ are the retarded and advanced
Greens functions, well known from equilibrium theory and carry
information about the spectrum of the system. In particular, one
obtains the spectral density simply from 
\begin{equation}
\hat A = i (\hat G^{\rm R} - \hat G^{\rm A}) \ .
\end{equation}
The spectral function generally obeys the normalization condition
\begin{equation}
\int_{-\infty}^\infty \frac {\upd \epsilon} {2\pi} \ \hat A(\vect k,
\epsilon, \vect r) = \hat 1 \,.
\end{equation}
The $\vect k$-integration of the spectral function yields the density
of states, here defined as number of states per unit energy and
volume:
\begin{equation}
\int \frac {\upd^3 k} {(2\pi)^3}\ \hat A(\vect k, \epsilon, \vect r) =
2\pi \hat \nu(\epsilon, \vect r) \ . 
\end{equation}
The lesser component $\Gless(\vect k, \epsilon, \vect r, t)$ describes
the occupation of states of the many particle system and is
given in terms of electron field operators by
\begin{equation}
\Gless_{\alpha,\beta}(\vect r, \vect r', \vect t, \vect t') = i\avg{\psi_\beta^\dagger(\vect r',\vect t') \psi_\alpha(\vect r,\vect t)} \ ,
\end{equation}
where the grandcanonical average is taken and the indices $\alpha,\beta$ are one of $\UPDOWN$.
In equilibrium, it takes the form
\begin{equation}
 \Gless(\vect k, \epsilon, \vect r) = i \hat A(\vect k, \epsilon, \vect r) f_{\rm D}(\epsilon)
\end{equation}
with the Fermi-Dirac distribution function
\begin{equation}
f_{\rm D}(\epsilon) = \frac 1 {e^{\beta (\epsilon-\eF)} + 1} \ , 
\end{equation}
where $\beta = 1/k_{\rm B}T$ denotes the inverse temperature.

From the lesser Greens function $\Gless$ we can easily obtain various
physical quantities of interest such as the quasi-particle spin-charge
density
\begin{equation}
  \label{eq:DensityFromGlesser}
  \hat N(\rt) = -e\int \frac {\upd^3 k} {(2\pi)^3} \int_{-\infty}^\infty
  \frac {\upd\epsilon}{2\pi i} \  
  \Gless(\vect k,\epsilon,\rt) \ ,
\end{equation}
and spin-charge current density
\begin{equation}
\label{eq:CurrentFromGlesser}
\vect {\hat J}(\rt) = -e\int 
\frac {\upd^3 k}{(2\pi)^3} \int_{-\infty}^\infty 
\frac {\upd\epsilon}{2\pi i} \ 
\vect v_k \Gless(\vect k,\epsilon,\rt) \ ,
\end{equation}
where the quasi-particle velocity is $\vect v_k = \partial_{\vect k}
\epsilon_{\vect k} = \vect k / m$.  The spin-charge density $\hat N$
is a $2\times2$ matrix in spin space and its trace yields the
charge-density, $n_c = \Tr\hat N$ while the spin-density is given by
$\vect s = -\frac{\hbar}{2e}\Tr(\sig \hat N)$. In a similar manner, we
obtain the charge current $\vect j_c = \Tr\ \vect{\hat J}$ and the
spin-k current $\vect j_{k}
= -\frac{\hbar}{2e}\Tr(\hat\sigma_k \vect{\hat J})$.

In order to find $\check G$ for a specific physical system, we need an
equation of motion, called the Dyson equation, which can be written in
the two forms
\begin{eqnarray}
  \label{eq:Dyson}
  \left[ (\epsilon \hat 1 - \hat \calH_0 - \hat \calH_{\rm S})\ \check
    1 - \check \Sigma \right] \otimes \check G = \check 1 \ ,
  \nonumber\\ 
  \check G \otimes \left[ (\epsilon \hat 1 - \hat \calH_0 - \hat
    \calH_{\rm S})\ \check 1 - \check \Sigma \right] = \check 1 \ . 
\end{eqnarray}
The self-energy $\hat \Sigma$ incorporates scattering by magnetic and
non-magnetic impurities. The leading contribution is calculated in the
self-consistent Born approximation, which truncates the series of
irreducible diagrams due to multiple impurity scattering after the first one:
\begin{equation}
  \label{eq:SelfEnergyDiagram}
  \includegraphics[height=1.6em]{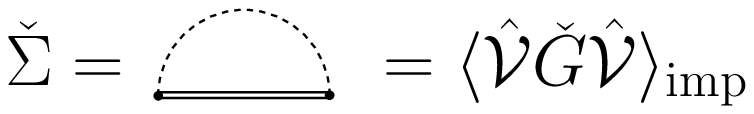}
\end{equation}
In physical terms, this means that all kinds of interference effects
like weak localization are dropped from the theory. In this
approximation, the self-energy for spin-independent impurity
scattering takes the form
\begin{eqnarray}
  \label{eq:SigmaImpurity}
  \check \Sigma_{\rm i} &=& \xii \int \frac {\upd^3 k} {(2\pi)^3}\
  \check G\,.
\end{eqnarray}
Similarly, for magnetic impurity scattering we obtain
\begin{eqnarray}
  \label{eq:SigmaMagImpurity}
  \check \Sigma_{\rm mag} &=& \sum_{i,j=1}^3 \xim_{ij} \int \frac {\upd^3
    k} {(2\pi)^3} \ \hat\sigma_i \check G \hat\sigma_j \,. 
\end{eqnarray}
The tensorial structure of $\ximTens$ accounts for situations in which
scattering is anisotropic in spin space. For example, if an exchange
coupling between the internal impurity spin and the ferromagnetic
order parameter exists, the spin is preferably aligned along this
direction. Consequently, the impurity will scatter the electron with
a magnitude depending on its spin. In the case of uniaxial
symmetry (the symmetry axis is denoted by $\vect n$),
\begin{equation}
  \label{eq:chim_ij}
   \xim_{ij} = \xims (\delta_{ij} - n_i n_j) + \ximp n_i n_j \ .
\end{equation}
For the above example, the unit vector $\vect n$ actually corresponds
to the local magnetization direction $\m$.  We assume throughout the
rest of this article that scattering is weak such that
\begin{equation}
  \label{eq:weak_scattering}
  \Sigma \ll \eF \ ,
\end{equation}
where $\eF$ is the Fermi energy of the conduction electrons.
Put in other words, this means that transport quantities like
density of states do vary very slowly on the scale of the relaxation
rates given by $\Sigma$.
This assumption will allow us later to make use of the quasi-particle
approximation.

\subsection{Spectral Density}
The first step in solving the Dyson equation (\ref{eq:Dyson}) is to
find a solution for the spectral function. To do so, we write equations
for the retarded/advanced components of $\check G$,
\begin{eqnarray}
  \label{eq:Dyson_ret_adv}
  \left([\epsilon - \epsilon_{\vect k} + \delmu(\vect r)] \hat 1 +
    \frac\Delta2 \m(\vect r) \sig\right) \otimes \hat G^{\rm R/A} =
  \hat 1 \,, \nonumber\\ 
  \hat G^{\rm R/A} \otimes \left([\epsilon - \epsilon_{\vect k} +
    \delmu(\vect r)] \hat 1 + \frac\Delta2 \m(\vect r) \sig\right)  =
  \hat 1 \,.
\end{eqnarray}
$\hat G^{\rm R/A}$ only differs by the boundary condition which can
be simply incorporated by substituting $\epsilon \rightarrow \epsilon \pm i\eta$
in Eqn.~(\ref{eq:Dyson_ret_adv}).
Furthermore, since $\Sigma \ll \eF$, we will not include impurity self
energy corrections to the spectrum, i.e. we neglect the line
broadening and assume a delta-peaked spectrum, commonly
referred to as the quasi-particle approximation.  Also, the external
electric field affects the spectrum only in second order of the field.

To proceed, we perform a gradient expansion
\begin{equation}
  \otimes \approx \mathbbm 1 + \frac i 2 ( \partialLeft{\vect
    r} \partialRight{\vect k}
  - \partialLeft{t} \partialRight{\epsilon} - \partialLeft{\vect
    k} \partialRight{\vect r}
  + \partialLeft{\epsilon} \partialRight{t} ) + \dots \ , 
\end{equation}
and determine $\hat G^{\rm R/A}$ iteratively by the order of the
gradient. It turns out, that we need only up to order $\partial_{\vect r}^2$.  For
example, one finds for the first two terms of the spectral density
$\hat A = \hat A_0 + \hat A_1 + \dots$.
\begin{widetext}
\begin{eqnarray}
  \label{eq:A_corrected}
  \hat A_0 &=& 2\pi \left[\PU \delta(\epsilon-\epsilon_{\vect k}^\UP)
    + \PD \delta(\epsilon-\epsilon_{\vect k}^\DOWN) \right] \ ,\\ 
  \hat A_1 &=& -\frac\pi2 \sig\left(\m\!\times\!(\vect
    v_k\dr)\m\right) \left( \de\left[\delta(\epsilon-\epsilon_{\vect
        k}^\UP) + \delta(\epsilon-\epsilon_{\vect k}^\DOWN) \right] -
    \frac{2}{\Delta}\left[\delta(\epsilon-\epsilon_{\vect k}^\UP) -
      \delta(\epsilon-\epsilon_{\vect k}^\DOWN) \right] \right) \ . 
\end{eqnarray}
\end{widetext}
Here, we defined the projectors in spin-space $\hat P_\UPDOWN = \frac12 (\hat
1 \pm \m\sig)$ that project on the spin up/down direction and
$\epsilon_{\vect k}^\UPDOWN \equiv \epsilon_{\vect k} \mp
\frac\Delta2$ is the dispersion relation for majority and minority
spin bands which are exchange split due to the {\it s-d} coupling.

The density of states for majority and minority spin bands are
\begin{equation}
  \nu_\UPDOWN(\epsilon) = \int \frac {\upd^3 k} {(2\pi)^3} \
  \delta(\epsilon - \epsilon_{\vect k}^\UPDOWN) \,.
\end{equation}
With $\nu_\UPDOWN$ without energy argument we denote the density of states at
the Fermi-level.  The matrix density of states without magnetization
gradient present is thus
\begin{equation}
  \hat \nu = \frac12(\hat1+\m\sig) \nu_\UP + \frac12(\hat1-\m\sig)
  \nu_\DOWN = \nu_0 (\hat1 + P\ \m\sig) \ , 
\end{equation}
where we have introduced the polarization of the Fermi surface 
\begin{equation}
  P = \frac{\nu_\UP - \nu_\DOWN}{\nu_\UP + \nu_\DOWN}
\end{equation}
and $\nu_0 = \frac12 (\nu_\UP + \nu_\DOWN)$ denotes the average
density of states.

The term $\delmu$ in Eqs.~(\ref{eq:Dyson_ret_adv}) constitutes an
additional chemical potential shift due to the magnetization gradient
and is determined by enforcing local charge neutrality. The total
quasi-particle density (\ref{eq:DensityFromGlesser}) is constant
throughout the ferromagnet, i.e. $\Tr{\hat N(\vect r)} = -e N_{\rm
  Ion}$, which leads to
\begin{equation}
  \label{eq:chemicalPotential}
  \delmu(\vect r) = \frac1{12}\left(1 - 2P \frac {2\eF}
    {\Delta}\right) \ \frac1{2M} (\dr\m)^2  \ . 
\end{equation}

\subsection{Kinetic Equation for the Non-Equilibrium Distribution}

To obtain a kinetic equation for the non-equilibrium distribution
function, we subtract the left- and right-conjugated Dyson equations
(\ref{eq:Dyson}),
\begin{equation}
  \label{eq:DysonSubtracted}
  \left[ (\epsilon \hat 1 - \hat \calH_0 - \hat \calH_{\rm S})\ \check
    1 - \check \Sigma \stackrel{\otimes}{,} \check G \right] = 0\,. 
\end{equation}
The equation for the lesser component is
\begin{widetext}
  \begin{equation}
    \label{eq:full_kinetic_eqn}
    -i\left[(\epsilon - \epsilon_{\vect k} + \delmu + e\varphi)\hat 1
      + \frac\Delta2 \m\sig \stackrel{\otimes}{,} \Gless \right] = 
    \frac 1 2 \left\{ \hat A \stackrel{\otimes}{,} \Sless \right\}
    - \frac 1 2 \left\{ \hat \Gamma \stackrel{\otimes}{,} \Gless \right\}
    -i\left[ \Re \hat \Sigma \stackrel{\otimes}{,} \Gless \right]
    +i\left[ \Re \hat G \stackrel{\otimes}{,} \Sless \right]
  \end{equation}
\end{widetext}
where the (anti-)commutators are defined by $\{\hat
A\stackrel{\otimes}{,} \hat B \} = \hat A\otimes \hat B + \hat
B\otimes\hat A$ and $[\hat A\stackrel{\otimes}{,} \hat B ] = \hat
A\otimes \hat B - \hat B\otimes\hat A$.  The imaginary part of the
self-energy
\begin{equation}
  \hat \Gamma = -2\Im \hat\Sigma^{\rm R} = i(\hat\Sigma^{\rm R} -
  \hat\Sigma^{\rm A}) 
\end{equation}
describes relaxation due to impurity scattering, while the real part
is responsible for changes in the energy dispersion relation.
There exists a Kramers-Kronig relation between real and
imaginary part,
\begin{eqnarray}
\label{eq:RealPartKramersKronig}
  \Re \hat \Sigma(\epsilon) = \frac12 (\hat \Sigma^{\rm R} + \hat
  \Sigma^{\rm A}) = 
  \frac 1 {2\pi} \calP\int_{-\infty}^\infty \upd \epsilon' \ 
  \frac {\hat \Gamma(\epsilon')} {\epsilon - \epsilon'} \ .
\end{eqnarray}
However, as can be seen from the integral representation in this formula,
the real part depends on the complete electronic spectrum of the system, 
since the scattering rate is directly related to the density of states,
$\Gamma(\epsilon) \propto \nu(\epsilon)$. Considering that the dynamics
accompanied by a rotation of the magnetization direction, be it in
time or space, is affecting only an energy region of the order of
$\Delta$, these changes constitute only a tiny fraction of the whole
energy range. Thus, corrections due to magnetization dynamics to the
real part can be neglected when compared to the whole background
contribution, which then is just a constant (however formally diverging
due to the assumption of $\vect k$-independent impurity scattering)
and merely renormalizes the electronic spectrum. In fact, the only reason
to include impurity scattering is to add momentum and spin relaxation to
the conduction electron system, as contained in the imaginary part
of the self-energy, $\hat\Gamma$.
Therefore, the two commutators involving real parts can
be dropped from equation (\ref{eq:full_kinetic_eqn}).

Generally, we can distinguish the contributions to the kinetic
equations emerging from two regions in energy. The electronic states deep
inside the Fermi sea are affected by an inhomogeneous exchange
splitting $\Delta\m(\vect r)$. The fraction of the Fermi sea that
contributes is given by $\Delta/\eF$ and not necessarily
small. However, due to the Pauli principle, these states are fully
occupied for all reasonable temperatures and the change in the
spectrum does not affect the dynamics of the mobile electrons close to
the Fermi surface. These are responsible for the dynamics in the
quantum kinetic equation, since the electrons in an energy window
given by the temperature, voltage or other low-energy scales have the
freedom to move. These differences can be used to eliminate the
high-energy contribution from our quantum kinetic equation. We
perform a series of three steps to obtain an equation
describing the low-energy dynamics alone.

First, we eliminate the electric potential by the substitution $\omega
= \epsilon + e\varphi$, which transforms the derivatives according to
\begin{equation}
  \partial_\epsilon \rightarrow \partial_\omega\ , \qquad \dr
  \rightarrow \dr - e\vect E \partial_\omega 
\end{equation}
with the electric field $\vect E = -\dr\varphi$.

Secondly, we make the ansatz 
\begin{equation}
  \label{eq:Gless_ansatz}
  \Gless(\vect k, \omega, \vect r) = i\hat A(\vect k, \omega, \vect
  r)\  f_{\rm D}(\omega - e\varphi) + \delta \hat G(\vect k, \omega,
  \vect r).
\end{equation}
The first term drops out from the kinetic equation
(\ref{eq:full_kinetic_eqn}), leaving an equation for $\delta \hat G$
alone.  $\delta \hat G$ has two very practical properties. It is
proportional to the electric field, which allows us to drop the term
$e\varphi(\vect r) \delta\hat G$, since we are interested only in
linear response to an external field. The external potential is
incorporated via appropriate boundary conditions that are concretized
below. Furthermore, $\delta \hat G$ is peaked around the Fermi-level,
reflecting the fact that non-equilibrium processes take place
only in the vicinity of $\eF$, provided the temperature is low enough.
In fact, we use the zero-temperature approximation $ \frac {\partial
  f_{\rm D}} {\partial \epsilon} = -\delta(\epsilon-\eF) $ throughout
this work.

This brings us directly to the third step which consists of
integrating the whole equation over energy after setting $\omega =
\eF$ in all prefactors to $\delta\hat G$ on the right-hand-side of
equation (\ref{eq:full_kinetic_eqn}). The spectral densities in the
collision integral become $\hat A(\vect k,\eF,\vect r)$, which means
we neglect the energy-dependence of the scattering rates.  This is
perfectly compatible with the linear response regime, since a more
in-depth investigation shows that corrections due to the energy
dependence of $\hat A$ are of quadratic order in the electric field.

For the stationary situation ($\partial_t = 0$), the resulting kinetic
equation for 
\begin{equation}
  \hat g(\vect k, \vect r) = \int \frac {\upd \omega}
  {2\pi i} \ \delta \hat G(\vect k, \omega, \vect r) 
\end{equation}
finally reads
\begin{equation}
  \label{eq:kineticEquationForg}
  \vect v_k \dr \hat g - i \frac \Delta 2 \left[ \m \sig
    \stackrel{\circ}{,} \hat g \right] + (\dr\delmu) \dk\hat g = \hat
  \calI_{\rm i}[\hat g] + \hat \calI_{\rm m}[\hat g]\ . 
\end{equation}
The collision integral takes the form
\begin{eqnarray}
  \label{eq:CollisionIntegral}
  \hat \calI_{\rm i}[\hat g]  & = &  \xii\int \frac {\upd^3 k'} {(2\pi)^3} \left[
    \frac 1 2  \left\{ \hat A(\vect k,\eF, \vect r)
      \stackrel{\circ}{,} \ \hat g(\vect k', \vect r) \right\}\right.
    \nonumber \\ & &  \qquad\left. - \frac
    1 2  \left\{ \hat A(\vect k',\eF, \vect r) \stackrel{\circ}{,} \
      \hat g(\vect k, \vect r) \right\} \right] 
\end{eqnarray}
and similar for the magnetic scattering contribution $\hat \calI_{\rm
  m}[\hat g]$. $\circ$ denotes the Wigner product
(\ref{eq:wigner_product}) with derivatives $\dr$ and $\dk$
only, and does {\it not} act on $\vect k'$. Again, we will need to
include only terms up to order $\dr^2$.  Equation
(\ref{eq:kineticEquationForg}) along with (\ref{eq:CollisionIntegral})
constitutes a linear integro-differential equation for $\hat g$ and
serves as the basis for our further analytical treatment.

Provided we have found a solution for $\hat g$ with appropriate
boundary conditions, we can finally determine the physical observables
of interest by simply substituting the ansatz (\ref{eq:Gless_ansatz})
into (\ref{eq:DensityFromGlesser}). In this way, we obtain the
quasi-particle spin-charge density
\begin{widetext}
  \begin{equation}
    \label{eq:FullDensity}
    \hat N(\vect r) =
    -e\int\frac{\upd\omega}{2\pi} \int \frac {\upd^3 k} {(2\pi)^3}
    \hat A(\vect k, \omega, \vect r)\ f_{\rm D}(\omega) 
   - e^2\varphi(\vect r) \frac1{2\pi}  \int \frac {\upd^3 k}
   {(2\pi)^3} \hat A(\vect k, \eF, \vect r) 
   + \hat n (\vect r)
 \end{equation}
\end{widetext}
with the low-energy spin-charge density excitations $\hat n = -e\int
\frac {\upd^3 k} {(2\pi)^3} \hat g$.  We made use of the
zero-temperature approximation and smallness of the external
potential in the linear regime, viz $f_{\rm D}(\omega-e\varphi) = f_{\rm D}(\omega) +
e\varphi\ \delta(\omega-\eF)$. The (spin-)current is obtained in a
similar straightforward manner.

On length scales much larger than the Thomas-Fermi screening length,
which is of the order of atomic distances in metallic materials,
local charge neutrality is fulfilled. Since the first term
on the right-hand side of (\ref{eq:FullDensity}) corresponds to
the equilibrium density which 
exactly neutralizes the material, the trace of the last two terms
has to vanish,
\begin{equation}
  \label{eq:ElectricPotentialFromN}
  \Tr\hat n(\vect r) = e^2 \varphi(\vect r) \frac1{2\pi} \int \frac
  {\upd^3 k} {(2\pi)^3} \Tr \hat A(\vect k, \eF, \vect r) \ .
\end{equation}
Therefore, $\Tr\hat n(\vect r)$ is directly related to the
local electric potential and thus the local electric field
$E(\vect r) = -\dr\varphi(\vect r)$ in the system.

Now we consider a region far from any inhomogenities in the magnetization
for which we want to specify an appropriate boundary condition.
In the absence of a magnetization gradient and non-equilibrium spin-excitations
in the conduction electron system, a less strict equality between the
last two terms of (\ref{eq:FullDensity}) holds,
\begin{equation}
  \label{eq:BoundaryConditionGeneral}
  \hat n(\vect r) = e^2 \varphi(\vect r) \frac1{2\pi} \int \frac
  {\upd^3 k} {(2\pi)^3} \hat A(\vect k, \eF, \vect r) \ .
\end{equation}
In a simple one-dimensional geometry and at the left and right
boundaries $x_{\rm L,R}$, the boundary condition simply becomes
\begin{eqnarray}
  \label{eq:BoundaryCondition}
  \hat n(x_{\rm L,R}) & = & e^2\varphi(x_{\rm L,R}) \left(\nu_\UP\PU +
    \nu_\DOWN\PD\right) \\\nonumber &=& e^2\varphi(x_{\rm L,R}) \nu_0
  \left(\hat1 + P 
    \m\sig \right) \ . 
\end{eqnarray}
Before continuing to solve the kinetic equation (\ref{eq:kineticEquationForg}) for $\hat
g$, let us specify a convenient geometry for the contact.

\section{Domain-Wall Resistance in Diffusive Wires with Current
  Perpendicular to Wall  (CPW) Geometry}

\newcommand{\gns} { \gamma_n^\perp }
\newcommand{\gjs} { \gamma_j^\perp }
\newcommand{\gnP} { \gamma_n^\UP }
\newcommand{\gjP} { \gamma_j^\UP }
\newcommand{\gnM} { \gamma_n^\DOWN }
\newcommand{\gjM} { \gamma_j^\DOWN }
\newcommand{\gnPM} { \gamma_n^\UPDOWN }
\newcommand{\gjPM} { \gamma_j^\UPDOWN }

\newcommand{\Del} { {\tilde\Delta} }

\newcommand{\px} { \partial }

\subsection{Model of the Contact}

We treat a ferromagnet in a quasi one-dimensional geometry such that
there are only gradients of the magnetization in the $x$ direction,
while in the $y$- and $z$-directions the system is homogeneous.
Furthermore, we assume a coplanar magnetization which allows
to parameterize the magnetization direction by a single angle
$\theta(x)$, defined by
\begin{equation}
  \m(x) = \left( \begin{array}{c}
      0 \\ \sin\theta(x) \\ \cos\theta(x)
    \end{array} \right) \ .
\end{equation}
Current flow perpendicular to the domain wall implies that the
direction of $\vect j$ is along the $x-$axis, so that the problem
becomes effectively one-dimensional.  Therefore, in this case, every
quantity of interest only depends on $x$, we can set $\partial_y
= \partial_z = 0$ and abbreviate $\partial \equiv \partial_x$.

\begin{figure}[t]
  \includegraphics[width=0.8\columnwidth]{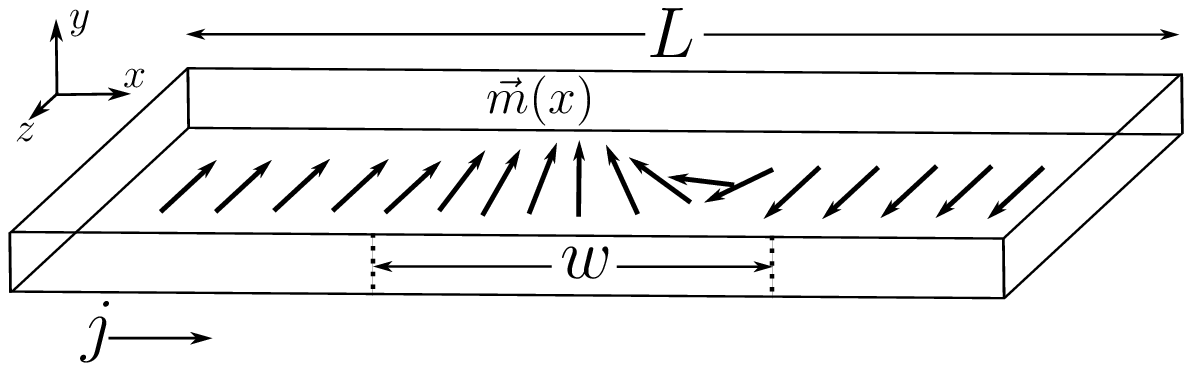}
  \includegraphics[width=0.7\columnwidth]{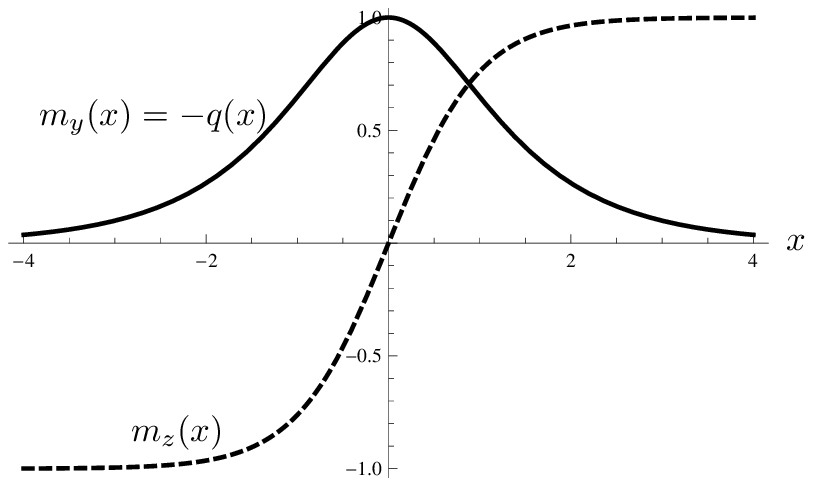}
  \caption{\label{fig:contact_with_domain_wall} Contact of length $L$ 
    with a Bloch domain wall of characteristic size $w$ situated in between.
    The domain wall is described by the magnetization gradient $q(x) = \px \theta(x)$.
    The lower plot shows $m_y(x) = -q(x)$ and $m_z(x)$ according to equation (\ref{eq:DomainWallProfile})
    with unity wall width $w=1$.}
\end{figure}

The contact of length $L$ contains a domain wall of characteristic
width $w$. The left- and rightmost parts have opposite magnetization
directions and a so far arbitrary magnetization profile
$\theta(x)$ connects the two magnetic domains. The situation is
sketched in Fig.~\ref{fig:contact_with_domain_wall}.

It is then convenient to perform a local SU(2) gauge transformation in
order to rotate the coordinate system used to represent the
spin-orientation such that its $z$-axis is always aligned to the local
magnetization direction $\m$. The unitary operator ($\hat {\mathcal
  U}^\dagger = \hat {\mathcal U}^{-1}$)
\begin{equation}
\label{eq:GaugeTransformation}
\hat {\mathcal U} = e^{-\frac i 2 \theta \hat \sigma_x}
\end{equation}
exactly performs this rotation in spin-space using the previously
introduced angle $\theta$, so that $\hat{\mathcal U}\ \m\sig \
\hat{\mathcal U}^{-1} = \hat \sigma_z$. As $\theta$
is spatially dependent, the derivatives obtain an additional term
after the transformation,
\begin{equation}
  \label{eq:GaugeDerivative}
  \hat{\mathcal U}\ \px \ \hat{\mathcal U}^{-1} = \hat 1 \px + \frac i
  2 q(x) \hat \sigma_x \equiv \hat \px \ .
\end{equation}
Henceforth, we have to deal with the gradient $q(x) = \px \theta$ in
our equations and which fully parameterizes the domain wall.
The gauge transformation (\ref{eq:GaugeTransformation}) rotates
the orthonormal basis system $(\vect v\!\times\!\m, \vect v, \m)$ such that
it becomes the canonical set basis vectors, $(\vect e_x, \vect e_y, \vect e_z)$.
The unit vector $\vect v$ points in the direction of change of the magnetization $\m$,
i.e. $\px \m = q \vect v$. Note that $\vect v$ is perpendicular to
$\m$, since $\m$ is a unit vector. 

An analytical form of the domain wall profile can be derived in the
context of minimising the free energy of the ferromagnetic material,
commonly a compromise between exchange and anisotropy
energy. Typically, one finds
\begin{equation}
  \label{eq:DomainWallProfile}
  \cos\theta(x) = \tanh(\tfrac x w) \ ,
\end{equation}
so that
\begin{equation}
  \label{eq:DomainWallProfileQ}
  q^2(x) = \frac 1 {w^2 \cosh^2( \tfrac x w)}
  = \frac 1 {w} \frac {\partial}{\partial x} \tanh(\tfrac x w) \ ,
\end{equation}
where $w$ constitutes a typical length scale over which the domain
wall extends.

\subsection{A Hierarchy of Equations}

\newcommand{\Gl}  {\myTensS G_{\rm ll}}
\newcommand{\Gt}  {\myTensS G_{\rm tt}}
\newcommand{\Gan} { \myTens\Gamma } 
\newcommand{\Gaj} { \myTens\Pi } 

\newcommand{\Vect}[1] { \vec {#1} }

In light of the solution procedure that is to come, it is convenient
to use a $4$-component vector representation (which is denoted by an arrow to
distringuish it from the 3-component real-space vectors set in boldface),
in which the spin-charge density excitations and current density take the form, respectively,
\begin{eqnarray*}
  \Vect n &=& (n_+,n_-,n_\UP,n_\DOWN) \ ,\nonumber\\
  \Vect j &=& (j_+,j_-,j_\UP,j_\DOWN) \ .
\end{eqnarray*}
Starting from the $2\!\times\!2$-matrix reprentation after the gauge transformation,
$\hat n = \frac {n_c}2 \hat 1 + n_x \hat\sigma_x + n_y \hat\sigma_y + n_z \hat\sigma_z$,
we define the spin-up/down densities $n_\UPDOWN = \frac {n_c}2 \pm n_z$.
Additionally, the transverse spin-degrees of freedom are transformed
according to $n_\pm = n_x \pm i n_y$, which corresponds to
circularly polarized transverse spin excitations. This basis is convenient because
it diagonalizes the equations of motion for the transverse dynamics in absence of
magnetization gradients, see Eqn. (\ref{eq:finalDGLforN}) and following.

The full kinetic equation for $\hat g$ is still very involved, so we
need to perform additional simplifying steps.  Therefore, we multiply
Eq.~(\ref{eq:kineticEquationForg}) with $v_x^n$ ($n \ge 0$) and afterwards
integrate over the whole $k$-space. This involves evaluating terms of
the form
\begin{widetext}
  \begin{eqnarray}
    \mv{v_x^n \partial_{k_x}^m \hat g} &=& \int \frac {\upd^3 k}
    {(2\pi)^3} v_x^n \partial_{k_x}^m \hat g \nonumber  
    = (-1)^m  \int \frac {\upd^3 k} {(2\pi)^3} \hat g \partial_{k_x}^m v_x^n
    = \left\{ \begin{array}{cl}
        \frac {(-1)^m} {M^m} \frac {n!} {(n-m)!} \hat g^{(n-m)} &
        \rm{if}\ n\geq m\\ 
        0 & \rm{if}\ n < m
      \end{array} \right.
    \label{eq:TakingGAverages}
\end{eqnarray}
\end{widetext}
and defining moments of the Greens function $\hat g$
\begin{equation}
  \hat g^{(n)}= -e \int \frac {\upd^3 k} {(2\pi)^3} v_x^n \hat g\,.
\end{equation}
The first two momenta are $\hat g^{(0)}=\hat n$ and $\hat g^{(1)}=\hat
j$, obviously.  The kinetic equation (\ref{eq:kineticEquationForg})
turns into an infinite hierarchy of equations relating these
moments. It is nevertheless possible to find an analytical solution to
these equations in form of a systematic series expansion in the
function $q(x)$. As we will show below, it is then possible to truncate
the hierarchy by setting $\hat g^{(5)} = 0$ in order to calculate the
domain-wall resistance. The first two of these equations are
reminiscent of spin-charge continuity and diffusion equations and take
the explicit form
\begin{eqnarray}
  \label{eq:SpinContinuityDiffusionEquations}
  \Gan\Vect n + \myTens\px \Vect j &=& O(q) \ , \\
  \Gaj \left( \Vect j + \myTens D\ \px \Vect n \right) &=& O(q) \ ,
\end{eqnarray}
where we defined the derivative $\myTens\px$, modified by the SU(2) gauge
transformation (see equation (\ref{eq:GaugeDerivative})),
\begin{equation}
  \myTens\px \equiv \myTens1\px + q\myTens M_x
  \equiv \myTens1\px + \frac i2 q 
  \left(\begin{array}{cccc}0&0&1&-1 \\
      0&0&-1&1 \\ 1&-1&0&0 \\ -1&1&0&0  
    \end{array}\right) \ .
\end{equation}
The right-hand-sides of these equations include important
contributions that are due to modification of various transport
properties in presence of a magnetization gradient like the change in
the density of states which enters the collision integrals. Thus, the
right-hand-side vanishes as $q\rightarrow0$. Their explicit form along
with the equations for $\hat g^{(2)}$, etc. are rather lengthy and not
needed for our following discussion. Further details can be found
in the appendix in equations (\ref{eq:HierarchyZero})-(\ref{eq:HierarchyOneAdjusted}).

\newcommand{\diag} {\mathrm{diag}}
The various scattering rates associated with momentum relaxation
$\Gaj$ and spin-flip/-dephasing $\Gan$ are most conveniently
expressed, introducing the parameters $\gamma, \kappa$ and $\xi$, in
the following way:
\begin{widetext}
  \begin{eqnarray}
    \nonumber
    \Gaj & = & \diag\left(\gjs+i\Delta,\gjs-i\Delta,\gjP,\gjM\right) =
       \gjs\diag\left(1+\frac i \chi,1 - \frac i \chi,1 + \gamma,1 - \gamma\right)
    \\\nonumber
    \Gan & = & \left(\begin{array}{cccc}
        \gns+i\Delta & 0 & 0 & 0 \\ 0 & \gns-i\Delta & 0 & 0 \\ 0 & 0 &
        \gnP & -\gnM \\ 0 & 0 & -\gnP & \gnM 
      \end{array}\right)
    = \gns \left(\begin{array}{cccc}
        1+\frac i {\xi \chi} & 0 & 0 & 0 \\ 0 & 1-\frac i {\xi \chi} & 0 & 0 \\
        0 & 0 & \frac12 (1-P)\kappa & -\frac12 (1+P) \kappa \\ 0 & 0 &
        -\frac12 (1-P) \kappa & \frac12 (1+P) \kappa 
      \end{array}\right)
  \end{eqnarray}
\end{widetext}
Spin precession is incorporated as well and manifests itself as an
imaginary part in the entries of the transverse subspace of $\Gan$ and
$\Gaj$.

In terms of the parameters of our specific model, the various
scattering parameters take the form
\begin{eqnarray}
  \chi &=& \frac{\gjs}{\Delta} = \frac{2\pi(\nu_\UP +
    \nu_\DOWN)}{2\Delta} \left(\xii + \ximp + 2\xims\right)
  \nonumber\\ 
  \xi &=& \frac{\gns}{\gjs} = 2 \ \frac{\ximp + \xims}{\xii + \ximp +
    2\xims} \ , \nonumber\\ 
  \gamma &=& P \ \frac{\xii + \ximp - 2\xims}{\xii + \ximp + 2\xims} =
  P \left( 1 - \xi\kappa \right) \ , \nonumber\\ 
  \kappa &=& \frac{2\xims}{\ximp + \xims} \ .
\label{eq:scatteringConsts}
\end{eqnarray}
$\chi$ measures the ratio of momentum relaxation rate to exchange
splitting $\Delta$, $\xi$ denotes the transverse spin-dephasing
strength and $\gamma$ can be considered as scattering asymmetry
parameter while $\kappa$ is the strength of spin-flip scattering.

Of course, the parameter $\gamma$ carries an implicit dependency on
the exchange splitting $\Delta$ entering through the polarization
parameter
\begin{equation}
  P(\Del) = \frac{\Del}{1+\sqrt{1-\Del^2}}\ ,
\end{equation}
expressed here by the dimensionless exchange splitting
$\Del\equiv\frac\Delta{2\eF}$.  Within our specific model of impurity
scattering, $\gamma = P(1-\xi\kappa)$ is not independent of
the other parameters. It is worth mentioning that this specific form
of $\gamma$ has to be employed in order to guarantee for a consistent
treatment of the gradient corrections in the collision integral.

Furthermore, the matrix of diffusion constants reads
\begin{eqnarray}
\label{eq:DiffusionMatrix}
\myTens D &=& \diag\left(\frac {\chi D_\perp} {\chi + i}, 
  \frac {\chi D_\perp} {\chi - i}, D_\UP, D_\DOWN\right) \nonumber\\
  &=& D_\perp \diag\left(\frac {\chi} {\chi + i}, \frac {\chi} {\chi - i},
  \frac{1+\Del}{1+\gamma}, \frac{1-\Del}{1-\gamma} \right) \ ,
\end{eqnarray}
where we introduced $D_\perp=\frac{2\eF}{3M} \frac1\gjs
= \frac13 \vF^2 \frac1\gjs$. Transverse spin excitations are also
subject to precession as they diffuse which results in the complex
values of the effective transverse diffusion constants.

Finally, let us stress that the reason for the inclusion of the
moments up to $\hat g^{(4)}$ lies in the fact that in our regime of
investigation the precession length $\lprec \equiv 2\pi\frac
{\vF}{\sqrt 3 \Delta}$ is, besides the Fermi-length, the smallest
length scale in the system. As can be seen from the equation
(\ref{eq:GreenTransversal}), $\lprec$ describes the period of
oscillation of transverse spin excitations.  In the diffusive
approximation, as for example used by Bergeret {\it et al}
\cite{Bergeret2002}, the scattering mean free path $\ls$ is the
smallest length scale in the system (besides the Fermi wave length)
and not $\lprec$. Hence, it is possible to truncate the hierarchy of
equations already by $\hat g^{(3)} = 0$, which results in only two
equations that are just the spin-chage continuity and diffusion
equation. The latter is obtained by plugging the equation for $\hat
g^{(2)}$ into the equation for $\hat g^{(1)}$ (see also appendix, for
more details).

\subsection{Solving the Hierarchy of Equations}
We eliminate the higher order moments $\hat g^{(n\geq1)}$ by
iteratively substituting the equations into each other, carefully
keeping terms that contribute up to order $q^2$.  We find a
differential equation for the vector of quasi-particle excitations
$\Vect n$ of the form
\begin{eqnarray}
  \label{eq:finalDGLforN}
  \left(\Gan - \myTens D \px^2\right) \Vect n &=& \myTens W(q) \Vect n 
\end{eqnarray}
where the differential operator $\myTens W(q)$ vanishes for
$q\rightarrow0$ and contains all possible terms up to order $q^2$. In
the homogeneous ($q=0$) and collinear case,
Eq.~(\ref{eq:finalDGLforN}) corresponds to the transport equation
used by Valet and Fert \cite{Valet1993}.  Let us stress that $\Vect n$
represents the unscreened spin-charge density excitations, while the
true, screened quantity $\Vect n^{(s)}$ is, according to
Eq.~(\ref{eq:FullDensity}),
\begin{equation}
  n^{(s)}_\UPDOWN(x) = n_\UPDOWN(x) - \nu_\UPDOWN(x) e^2 \varphi(x) \ .
\end{equation}
Here, spin-excitations are not screened since our model does not
include a spin-dependent interaction.

Generally, $\myTens W(q)$ contains terms of the form $\myTens Y_{ijk}
\px^i q \px^j q \px^k$, $\myTens Y_{ij} \px^i q \px^j$ and $\myTens
Y_{i} \px^i$, where the order of $q$ and $\px$ is crucial, since $\px$
acts on everything to its right. The constant matrices $\myTens Y$ that
depend on our set of parameters, $\Del, \gamma, \xi, \kappa$ can
be obtained in a straightforward manner from equations
(\ref{eq:HierarchyZero})-(\ref{eq:HierarchyFour}) by collecting
all terms associated with the corresponding factor $\px^i q \px^j q \px^k$.
Restricted to terms that contribute to DWR, its explicit form (see eqn.
(\ref{eq:WDefinition})) is given in the appendix.

Since we have a perturbative treatment in $q$ in
mind, we determine the Greens function of Eq.~(\ref{eq:finalDGLforN})
\begin{equation}
\label{eq:DefGreensFunction}
\left(\Gan - \myTens D \px^2\right) \myTens G(x) = \myTens 1 \delta(x) \ .
\end{equation}
Separated into longitudinal (l) and transversal (t) subspace,
the Greens function is
\begin{eqnarray}
  \myTens G(x) &=& \left( \begin{array}{cc} \Gt(x) & 0 \\ 0 &
      \Gl(x)  \end{array} \right) \ , \nonumber
  \\ \label{eq:GreenLongitudinal}
  \Gl(x) &=& \myTensS H e^{-\frac{\abs{x}}\lambda} + \myTensS K \frac
  {\abs{x}} \lambda \ , \\ 
  \label{eq:GreenTransversal}
  \Gt(x) &=& \left( \begin{array}{cc} f(x) & 0 \\ 0 &
      f^*(x)  \end{array} \right) \ ,
  \\\nonumber
  f(x) & = & \frac{\chi+i}{\chi D_\perp}\ \frac i{2k} e^{i
    k\abs{x}} \ ,
  \\\nonumber
  k^2 & = & \left(\frac{2\pi}{\lprec}\right)^2 
  \left(1-i\chi\right)\left(1-i\chi\xi\right) \ . 
\end{eqnarray}
Matrices in the $2\!\times\!2$ subspaces are denoted by a single
underbar as compared to the double underbar which indicates a
$4\!\times\!4$ matrix. The index l and t refers to longitudinal and
transverse componenents, respectively, so that for example
\begin{equation*}
  \myTens W = \left( \begin{array}{cc} \myTensS W_{\rm tt} & \myTensS
      W_{\rm tl} \\ \myTensS W_{\rm lt} &  \myTensS W_{\rm
        ll} \end{array} \right) \ . 
\end{equation*}

The longitudinal component $\Gl$ consists of two contributions. The
first term of $\Gl$ decribes spatial damping of spin-up/down non-equilibrium
excitations which manifests itself in the characteristic exponential
decay on the spin-diffusion length,
\begin{eqnarray}
\label{eq:SpinDiffusionLength}
\frac1{\lambda^2} \equiv \frac{\gnP}{D_\UP}+\frac{\gnM}{D_\DOWN} \ .
\end{eqnarray}
The second term of $\Gl$ describes the linear behaviour of the chemical potential in
a homogenous system in the absence of a magnetization gradient.
The two tensors
\begin{eqnarray}
  \myTensS H &=& \frac12 \frac{\lambda^3}{(D_\UP D_\DOWN)^2}\left(
    \begin{array}{cc}
      D_\DOWN^2\gnP & -D_\DOWN D_\UP \gnM \\ -D_\DOWN D_\UP \gnP &
      D_\UP^2 \gnM  \end{array}\right) \\
  \myTensS K &=& -\frac12 \frac{\lambda^3}{D_\UP D_\DOWN}
  \left(\begin{array}{cc}  \gnM & \gnM \\  \gnP & \gnP
    \end{array}\right) 
\end{eqnarray}
obey the useful identities
\begin{eqnarray}
  \myTensS\Gamma_{\rm ll} \myTensS K &=& 0 \ , \\
  \left(\myTensS\Gamma_{\rm ll} - \frac{1}{\lambda^2} \myTensS D_{\rm 
      ll} \right) \myTensS H &=& 0 \ , \\ 
  \frac 2 \lambda \myTensS D_{\rm ll} \left(\myTensS H - \myTensS
    K\right) &=& \myTens 1 \ ,
\end{eqnarray}
which can be invoked to easily verify that the longitudinal Greens
function $\Gl(x)$ in fact fulfills equation
(\ref{eq:DefGreensFunction}). The real part of the complex wave-vector
$k$ describes the precession of transverse non-equilibrium spin
excitations while its imaginary part is the damping due to dephasing
mechanisms. Therefore, the root of $k$ has to be chosen such that is
has a positive imaginary part.

In the following, we restrict ourselves to the regime $\chi \ll 1$,
viz. a momentum relaxation rate much smaller than the exchange splitting.
Again, since leading order correction term turns out to be of order $q^2$, we
drop any terms of higher order than that.
Later, we will see that this restricts the validity of
the result for the DWR to domain wall lengths larger than the
spin-precession length $\lprec$.

In this regime, the transverse oscillations are very rapid on the
scale of the magnetization gradient. Therefore, it is suitable to
eliminate the transverse degrees of freedom by first splitting the
equation of motion (\ref{eq:finalDGLforN}) for $\Vect n$ into
transverse and longitudinal parts,
\begin{eqnarray}
  \left(\Gan - \myTens D \px^2\right)_{\rm tt} \Vect n_{\rm t} &=&
  \myTensS W_{\rm tl} \Vect n_{\rm l} + \myTensS W_{\rm tt} \Vect
  n_{\rm t} \\ 
  \left(\Gan - \myTens D \px^2\right)_{\rm ll} \Vect n_{\rm l} &=&
  \myTensS W_{\rm ll} \Vect n_{\rm l} + \myTensS W_{\rm lt} \Vect
  n_{\rm t} 
\end{eqnarray}
and writing down the formal solution for the transverse component
\begin{equation}
  \Vect n_{\rm t}(x) = \int_{-\infty}^\infty \upd x'\Gt(x-x') \left[
    \myTensS W_{\rm tl} \Vect n_{\rm l}(x') + \myTensS W_{\rm tt}
    \Vect n_{\rm t}(x') \right] \ . 
\end{equation}
In the limit $\chi \ll 1$, $\Gt(x')$ varies on a scale determined by
$\frac{2\pi}k = \lprec$ which is much smaller than other length
scales of interest, which are variation of $\m$ and the external
electric field. In particular, for the length of the domain wall, $w
\gg \lprec$. Thus, to leading order in $\chi$, we can consider $\Gt$
as a representation of the Dirac $\delta$-function and perform the
integration. We obtain
\begin{equation}
  \label{eq:transverseSolution}
  \Vect n_{\rm t}(x) = \myTensS \calF \left[
    \myTensS W_{\rm tl} \Vect n_{\rm l}(x) + \myTensS W_{\rm tt} \Vect
    n_{\rm t}(x) \right] \ . 
\end{equation}
Here we introduced the spatially integrated transverse Greens
function,
\begin{equation}
  \myTensS \calF = \int_{-\infty}^\infty d x\ \Gt(x) = \left(\begin{array}{cc}  F & 0 \\ 0 &
      F^* \end{array}\right)\,,
\end{equation}
where
\begin{equation}
  F = \frac1{\gns} \frac{\chi\xi}{\chi\xi+i} \approx -i \frac1\gns
  \chi\xi = -i\frac1{\Delta} \ .  
\end{equation}
Note that $\myTensS \calF$ simply corresponds to the inverse of the
transverse part of $\Gan$, a fact which becomes clear by noting that
our approximation corresponds to the neglect of the transverse
diffusion term.

Additionally, we only need to keep the first term of
(\ref{eq:transverseSolution}) since the backaction on the transverse
dynamics, represented by the second term, appears only in higher
orders in $q$ and $\chi$. Explicitly, this is expressed by the fact
that to leading order in $q$, $\myTens W_{\rm tt}$ vanishes, so that
\begin{eqnarray}
  \label{eq:nTransverse}
  \Vect n_{\rm t}(x) & = & \myTensS \calF\ \myTensS W_{\rm tl} \Vect
  n_{\rm l}(x) + \myTensS \calF\ \myTensS W_{\rm tt} \myTensS \calF\
  \myTensS W_{\rm tl} \Vect n_{\rm l}(x) + \dots \nonumber \\  
  & = & \myTensS \calF\
  \myTensS W_{\rm tl} \Vect n_{\rm l}(x) + O(\chi^3) 
\end{eqnarray}
Putting this result back into the equation for the longitudinal
dynamics yields the formal solution,
\begin{eqnarray}
  \label{eq:formalSolutionlongitudinal}
  \Vect n_{\rm l}(x) & = &  \int_{-\infty}^\infty \upd x' \ \Gl(x'-x)
  \\\nonumber &&
  \times \left[\myTensS W_{\rm ll} \Vect n_{\rm l}(x') + \myTensS W_{\rm lt}
    \myTensS \calF\ \myTensS W_{\rm tl} \Vect n_{\rm l}(x') 
  \right] \ .
\end{eqnarray}

The boundary condition (\ref{eq:BoundaryCondition}) for the left and
right side of the contact reads
\begin{equation}
  \Vect n(\pm \tfrac L 2) = e^2\varphi\!\left(\pm\tfrac L 2\right) \
  \left( 0, 0, \nu_\UP, \nu_\DOWN \right) \ . 
\end{equation}
The zeroth order solution with externally applied bias voltage $V$,
that satisfies this boundary condition, is simply found to be
\begin{equation}
\label{eq:n0Solution}
\Vect n^{(0)}(x) = -  e^2 E x \  \left( 0, 0, \nu_\UP, \nu_\DOWN \right) \ ,
\end{equation}
where $E=\frac V L$ is the constant external electric field in absence
of the wall.

Substituting $\Vect n^{(0)}(x)$ into the right-hand-side of the
solution (\ref{eq:formalSolutionlongitudinal}) yields the second order
correction in $q$, $\Vect n^{(2)}(x)$. Due to charge-current
conservation, the current flowing through the contact is still
determined by $\Vect n^{(0)}$ because $\px\Vect n^{(2)}(x)$ is taken to
vanish at the boundaries by assuming that the contact is long enough
for any finite size effects to become negligible, i.e. $L\gg\lambda$.
In this regime, the exponential term in the
longitudinal Greens function (\ref{eq:GreenLongitudinal}) can be
neglected which means that spin-accumulation has faded near the reservoirs.
Then, the current can be deduced directly from
Eqs. (\ref{eq:n0Solution}), unaffected by the correction $\Vect
n^{(2)}$. Explicitly, this current reads
\begin{equation}
  \Vect j^{(0)} = -\myTens D \px \Vect n^{(0)} 
  = \left( 0, 0, \sigma_\UP, \sigma_\DOWN \right) E \ , 
\end{equation}
where the spin-resolved Drude conductivity of majority and minority spin channels is
given as usually by $\sigma_\UPDOWN = e^2 \nu_{\UPDOWN} D_{\UPDOWN}$.

However, $\Vect n^{(2)}(x)$ has an additional potential drop that
is extracted from the asymptotic behavior and that stems from the
second term in $\Gl(x)$,
\begin{eqnarray}
  \nu_0 e^2\delta V & = & n^{(2)}_{\rm c}(x\rightarrow+\infty) -
  n^{(2)}_{\rm c}(x\rightarrow-\infty) \\\nonumber &=& 2 n^{(2)}_{\rm
    c}(x\rightarrow\infty)\ . 
\end{eqnarray}
Here, we used that the charge component $n_c = n_\UP + n_\DOWN$ and
the second equality is due to symmetry of the contact.  Keeping the
current constant, the presence of the wall implies a change in the
externally applied potential, which directly translates into a
relative change in resistance. Hence, we define the DWR
\begin{equation}
  \label{eq:DWRGeneralFormula}
  \delta \rho_{\rm DW} \equiv \frac {\rho_{\rm DW} - \rho_0} {\rho_0}
  = \frac {\delta V} {V}  = \frac {2\ n^{(2)}_{\rm
      c}(x\rightarrow\infty)}{n^{(0)}_{\rm c}(+L/2)-n^{(0)}_{\rm
      c}(-L/2)} \ . 
\end{equation}

Of course, due to the perturbative nature of our treatment, the
correction $n^{(2)}$ has to be always smaller than $n^{(0)}$.  Since
it turns out that
\begin{equation}
  \Vect n^{(2)}(x\rightarrow\pm\infty) = \delta \rho_{\rm DW} \ \Vect
  n^{(0)}(\pm \tfrac L 2) 
\end{equation}
this condition is equivalent to $\delta \rho_{\rm DW} \ll 1$ which is
always the case.

\subsection{Results}
The correction to domain-wall resistance up to order $q^2 = (\px
\m(x))^2$ takes the form
\begin{equation}
\label{eq:DWRFinalResult}
\delta \rho_{\rm DW} \equiv \frac {\rho_{\rm DW} - \rho_0} {\rho_0}
 = \frac13 \frac {E_{\rm DW}}{\eF} \  f(\Del,\gamma,\xi,\kappa)
\end{equation}
where analogously to Ref.~[\onlinecite{Brataas1999}], we use the
domain wall energy defined by
\begin{equation}
  E_{\rm DW} = \frac {\hbar^2} {2M} \ 
  \frac 1 L \int^{+L/2}_{-L/2} q^2(x) \upd x = 
  \frac {\hbar^2} {2M} \ \frac C {wL} \ .
\end{equation}
The constant $C$ of order unity depends on the specific form of the
wall and we find the geometric scaling to be $1/wL$. For the domain
wall profile of eq. (\ref{eq:DomainWallProfileQ}) we obtain $C = 2$.
The scaling with $1/wL$ can be easily understood by realising that
corrections to the resistance arising from a gradient $q$ yield the
behavior $\delta \rho_{\rm DW} \propto q^2 \propto \frac1{w^2}$. Due
to physical reasons there cannot be a corrections linear in $q$
since the result should not depend on the sign of $q$, i.e. the sense
of rotation of $\m$. Since the total length of the contact is $L$ and
the domain wall constitutes only a fraction $w/L$, the correction for
the whole contact should indeed be $\delta g \propto 1/{wL}$.

A thorough investigation of the whole hierarchy of equations reveals
that the result obtained for $\delta \rho_{\rm DW}$ is valid for wall
lengths much larger than the spin-precession length, $w \gg \lprec$.
The mathematical reason for this condition is, that, even
though in the vicinity of the domain wall, the detailed profile of the
quasi-particle excitations $n^{(2)}(x)$ depends on the whole hierarchy
of equations (unless $w \gg \lsd$, with the transverse spin-diffusion
length $\lsd^2 \equiv D_\perp/\gns$), the asymptotic behavior of the
correction $n^{(2)}(x)$, is not affected by higher order contributions
of the multipole expansion. And in situations, in which finite size
effects from the contact geometry are negligible, $\delta \rho_{\rm
  DW}$ is solely determined by these asymptotics. In short, to obtain
the asymptotics, and thus the desired result, valid up to order $q^2$,
we need to take into account contributions up to $\Vect g^{(4)}$.

Up to order $\chi^2$, the final expression for $f$ can be split into
three parts,
\begin{eqnarray}
  f\left(\Del,\gamma,\xi,\kappa\right) = 
  \frac {f_0(\Del,\gamma) + \xi f_\xi(\Del,\gamma) + 
    \kappa\xi f_\kappa(\Del,\gamma)} {\Del^2 \left(1 - \gamma^2\right) 
    (1 + P(\Del-\gamma) - \gamma\Del)}\ ,
\end{eqnarray}
that take the explicit form
\begin{widetext}
\begin{eqnarray}
  f_0(\Del,\gamma) & = & 
  \tfrac {5\left(2(\Del-2P)+\Del^2(\Del-P)\right) + 
    \Del\gamma\left(13\Del\left(\gamma\Del-\gamma^2-1\right)-10\Del+8\gamma\right)-P\Del\gamma \left(19\left(\gamma\Del-1-\gamma^2\Del^2\right)+9\left(\Del^2-3\right)+4\gamma^2\left(1+2\Del^2\right)\right)}
{10 \Del} \nonumber\\
f_\xi(\Del,\gamma)&=&\tfrac {-\left(1-\gamma^2\right) \Del \left(1-\Del^2\right) + 2P(1 - \gamma\Del) \left(1-2\gamma  \Del +\Del^2\right)}
{\Del } \nonumber\\
f_\kappa(\Del,\gamma)&=&\tfrac {\left(\Del - \gamma(2-\gamma  \Del)\right) \left(8P - 4\Del + P\left(-7+P^2\right) \Del^2 + 2\Del^3\right)} {4 (1-\Del^2)}
 \ .
\end{eqnarray}
\end{widetext}
Let us reconsider the assumptions made during the derivation of result
(\ref{eq:DWRFinalResult}), $\chi \ll 1$ and $w \gg \lprec$ along with
(\ref{eq:conditionForMagnetization}). The former can be rewritten as
$\ls \gg \lprec$ where $\ls$ denotes the scattering mean free
path. This in fact implies that $\lprec$ is, besides the Fermi-length,
the smallest length scale in the system. Note that no assumption was
made on the relation between $\ls$ and the length of the domain wall.
However, we made the assumption of a diffusive contact which implies
that $L \gg \ls$.

For completeness, let us also specify the leading order correction to
the longitudinal current component,
\begin{eqnarray}
\Vect j^{(2)}_{\rm l} &=& -e  \ \left(j^{(2)}_{\rm s}, -j^{(2)}_{\rm s}\right) \\
j^{(2)}_{\rm s}(x) &=& e E D_\perp \nu_0 \frac13 \frac {E_q(x)}{\eF} \ g(\Del,\gamma,\xi,\kappa) \nonumber
\end{eqnarray}
which, in contrast to $\delta \rho_{\rm DW}$, is only valid for $w \gg
\lsd$. Obviously, due to conservation of charge current, this
correction constitutes a pure spin-current, $j^{(2)}_{\rm s}(x)$. The
$x$-dependence is inherited trivially from $E_q(x) = \frac
{q^2(x)}{2M}$.

\subsubsection{Quasiclassical Regime $\Del \ll 1$}
Let us first investigate the limit $\Delta \ll \eF$, commonly refered
to as the quasiclassical regime. However, due to the restrictions
imposed upon $\lprec$ and discussed above, $\Del$ cannot become
arbitrarily small.  In this regime, corrections to the electron
density of states play no role and the density of states can be
considered constant. Also, it turns out that gradient corrections to
the self-energies play no role so that only the contribution from
impurity scattering remains. We find in this limiting case
\begin{equation}
  \label{eq:DWRQuasiclassical}
  f\left(\Del,\gamma,\xi,\kappa\right) = \frac{\gamma^2}{1-\gamma^2}
  \left(\frac45+\xi\right) \frac 1 {\Del^2} \ , 
\end{equation}
with the scattering asymmetry parameter $\gamma$ as well as
spin-dephasing strength $\xi$.  This result shows the strong
enhancement by scattering asymmetry as already noted in previous works
\cite{Brataas1999}. It also displays the same $\Del$-dependency
already found in various other works.  Clearly, this result is due to
different conductivities in the two spin channels which are mixed in
regions of non-vanishing magnetization gradient $q$. A spin-up
electron incident on a domain wall attains a spin-down component since
electron spin direction does not instantaneously follow the local
magnetization direction $\m$ (known as spin mistracking). Since the
electronic spectrum plays no role in this limit, any asymmetry in the
conductivity of the two channels is due to the scattering asymmetry
$\gamma$, thus, there is no DWR as $\gamma\rightarrow0$.

Comparing this result with the works
\cite{Tatara1997,Brataas1999,Gorkom1999}, we find in this limit,
besides $\xi$, a different numerical prefactor. Adopting to our
notation, they have $ f = \frac95 \frac{\gamma^2}{1-\gamma^2} \frac 1
{\Del^2}$. To fathom this discrepancy, we stress that the specific
form of the longitudinal Greens function $\Gl(x)$ is a result of the
presence of spin-flip processes. In absence on these processes (which
is the case in aforementioned works), the spin-diffusion length
diverges, so that properly performing this limit yields the
longitudinal Greens function $\Gl(x) = \myTens H
e^{-\frac{\abs{x}}\lambda} + \myTens K \frac {\abs{x}} \lambda
\rightarrow (\myTens K - \myTens H) \frac {\abs{x}} \lambda$, which
produces qualitatively different results. This additional contribution
persists arbitrarily far away from the domain wall and leads to a
different result for $\delta \rho_{\rm DW}$. Calculating then the DWR
in the quasiclassical limit, we find that it is still smaller by a factor
of $2$ as compared to the result in \cite{Brataas1999}.
Nevertheless, this shows that spin-flip processes
are crucial and cannot be ignored, since the absence of the latter
leads to spin accumulation that does not decay even infinitely far
away from the domain wall and thus yields an additional contribution
to the DWR.

Concerning the work of \cite{Levy1997}, the main critic has been
mentioned in the introduction. Even restricting ourselves to the
quasiclassical regime, the use of a system consisting of an infinite
spin-spiral the inclusion of only up to p-wave component and the lack
of terms due to gauge transformation can be invoked to explain the
discrepancy to our fully microscopic results.

There is still the question about how small $\Del$ can become, since,
clearly, the limiting factor appearing in the expressions is
$1/\Del^{2}$. However since, within our specific model,
$$\gamma\propto P(\Del) = \frac{\Del}{2}+O[\Del]^3 \ , $$ 
there is no problem concerning $\delta \rho_{\rm DW}$, because then
$\Del^2$ cancels out. For the current, only $\Del$ cancels, but there
is another constriction leading to $E_q \ll \Delta$ which again stems
from the requirement that $\lprec$ is smaller than the scale of the
magnetization gradient. Hence again, we will not get into
trouble. This remains a problem in the work of Brataas and coworkers
\cite{Brataas1999}, since there, the impurity scattering times are
introduced phenomenologically and thus, in their scenario no
restriction is placed upon $\gamma$.

\subsubsection{Arbitrary $\Del < 1$}
Let us now have a closer look at the behavior in the whole range of
valid values for $\Del$. In this regime, the gradient corrections in
the collision integral become important and so is the influence of the
magnetization gradient on the electronic structure, viz., the density of
states.

In the half metallic limit the spin-flip length becomes arbitrarily
small, since $\lambda \rightarrow 0$ as $\Del \rightarrow 1$. Writing
$\Delta = \eF - \epsilon^2$ and owing to the condition that no length
should exceed the precession length $\lprec \ll \lambda$, we obtain
the requirement that $\frac {\epsilon^2} {\kappa(1-\gamma)} \gg
\gns\gjs $. However, this is not a big restriction considering our
assumption that scattering is weak, so that $\gjs,\gns \ll \eF$.

\begin{figure}[t]
\includegraphics[width=\columnwidth]{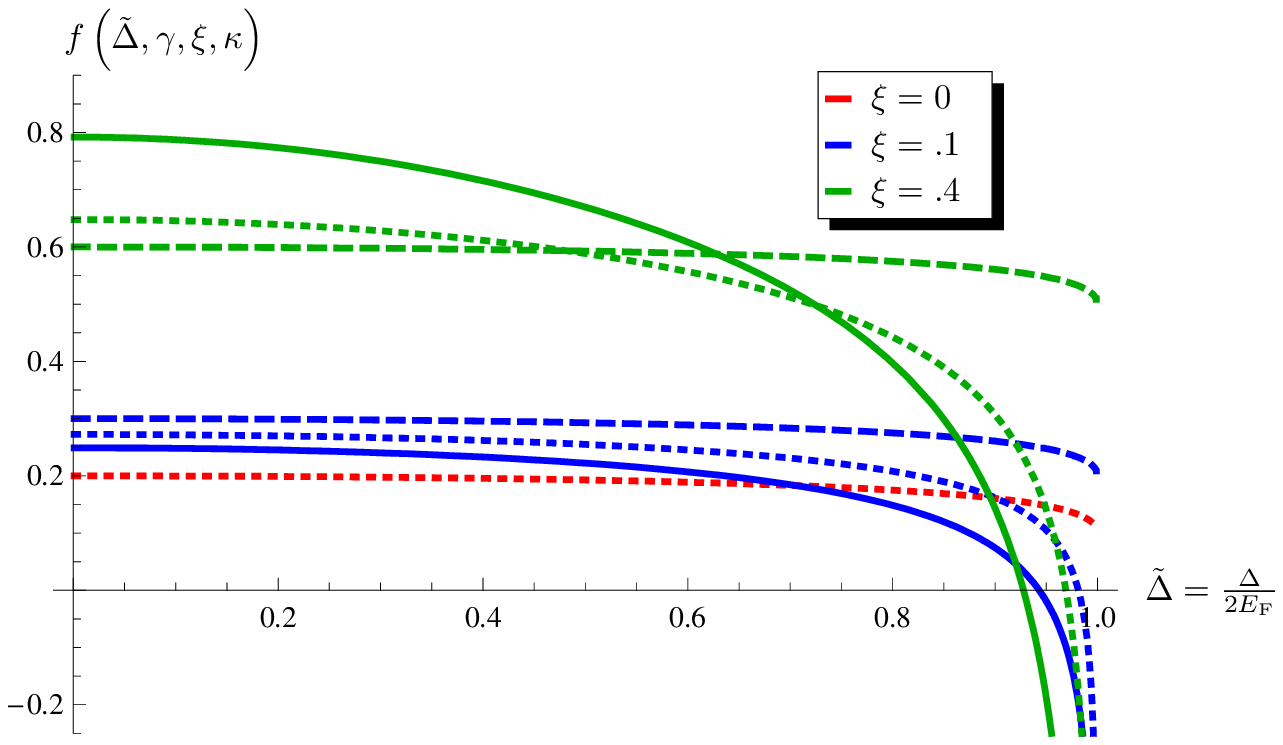}
\caption{\label{fig:DWRFull} Domain-wall resistance $\delta \rho_{\rm
    DW} = \frac13 \frac {E_{\rm DW}}{\eF}\ f(\Del,\gamma,\xi,\kappa)$
  as a function of exchange splitting $\Delta$ for various values of 
  $\xi\equiv\frac{\gns}{\gjs}$, the ratio of spin-dephasing rate and
  momentum relaxation rate and denotes the transverse spin-dephasing strength.
  $E_{\rm DW} = \frac {\hbar^2} {2M} \ \frac 2 {wL}$ is the domain wall energy.
  $\kappa=1$ implies spin-isotropic magnetic scattering of
  strength $\ximp=\xims$ (dotted lines), while $\kappa=2$ means
  $\ximp=0$ (normal lines) and $\kappa=0$ is the case of $\xims = 0$
  (dashed lines). $\gamma=P(1-\xi\kappa)$ is the scattering asymmetry
  parameter and is not an independant parameter in our model. }
\end{figure}

\begin{figure}[t]
\includegraphics[width=\columnwidth]{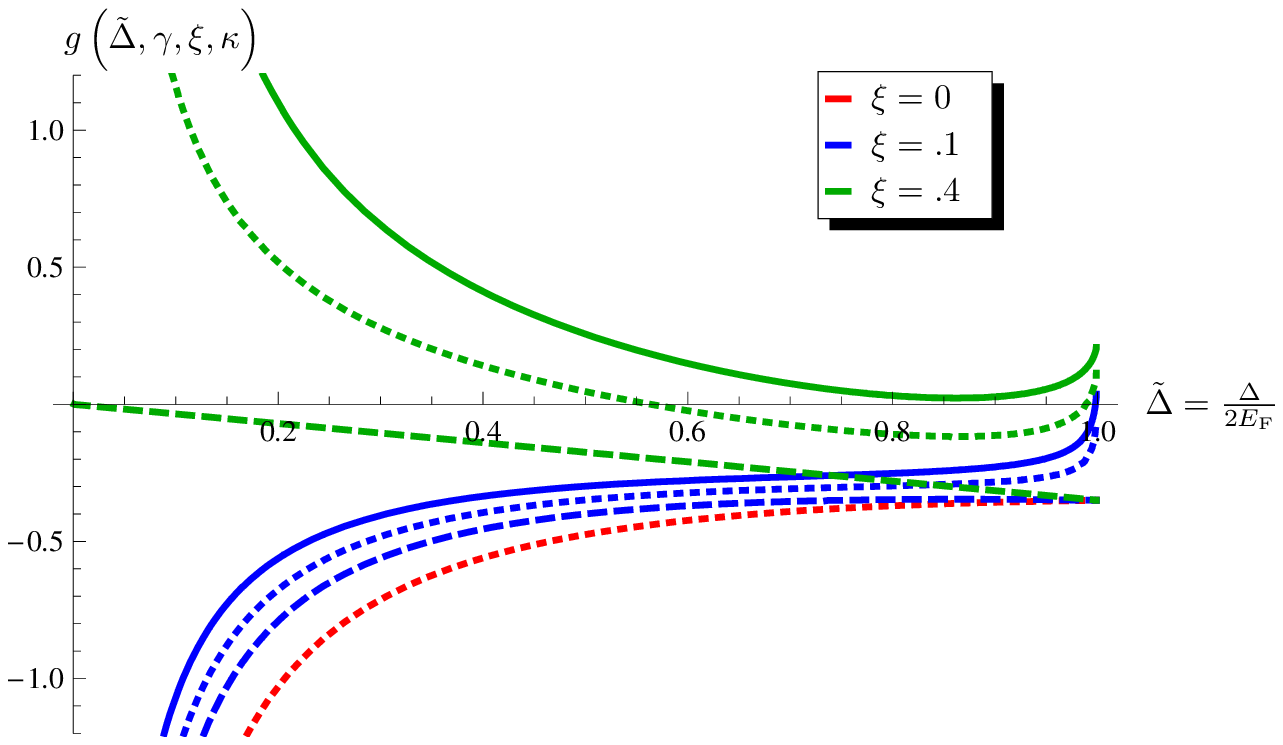}
\caption{\label{fig:CurrentFull} Additional
  magnetization-gradient-induced flow of longitudinal spin-current
  inside the domain wall where $E_q(x)= \frac {q^2(x)}{2M}$ does not
  vanish, so that $j^{(2)}_{\rm s}(x) = eE D_\perp \nu_0 \ \frac13
  \frac {E_q(x)}{\eF}\ g(\Del,\gamma,\xi,\kappa)$. Note that the
  divergence for $\Del\rightarrow0$ poses no problem, as discussed
  above. The magnetic scattering parameters are identical to
  FIG. \ref{fig:DWRFull}.}
\end{figure}

Note that within our model of impurity scattering we have only two
independent parameters. A convenient choice is to vary $\xi$ and
$\kappa$ independently, so that $\gamma = P(1-\xi\kappa)$ is
fixed. $\kappa$ can be also regarded as magnetic scattering asymmetry
parameter since we have $\kappa=0$ for $\xims=0$, $\kappa=1$ for
$\ximp=\xims$ and $\kappa=2$ for $\ximp=0$.  For various values of
$\xi$ and $\kappa$, the DWR is shown in figure \ref{fig:DWRFull} and
the corresponding change of the spin-current in the contact is depicted in
Fig.~\ref{fig:CurrentFull}.

We can see that the DWR does not vary strongly with $\Delta$ when this
parameter is small and decreases monotonously as $\Delta$ increases
even to the point where the DWR can become negative as one approaches
the half metallic regime. The latter happens only if $\xims\neq0$,
i.e. in the case of non-vanishing spin-flip scattering between the two
bands. Also, the magnetization-gradient-induced longitudinal
spin-current (c.f.~Fig.~\ref{fig:CurrentFull}) displays a qualitative
difference between the cases where $\xims$ vanishes an where it does
not. This clearly is a band structure effect and we find that this
also requires to include corrections to the collision integral. It
turns out that the presence of a magnetization gradient modifies the
density of states so that there are corrections of order $q^2$ to the
scattering rates that decrease the momentum relaxation rate and thus,
a reduction in resistivity. In fact, the density of states for the
minority spin channel strongly decreases as one approaches $\Del
\rightarrow 1$. Furthermore, we find that the neglect of said
corrections leads to a monotonic increase of the DWR with increasing
$\Delta$ as opposed to the present result. As a side note, we remark
that it is important to include corrections to the collision integral
up to the same order as in the approximation of the transport part of
the kinetic equation for the calculation to be consistent. Finally,
predictions for a given material would require realistic band
structure calculations, but the present calculations demonstrates the
possibility to have a negative DWR.

Let us finally compare our approach to the one of Brataas {\it et al}
\cite{Brataas1999} by temporarily excluding spin-flip processes. In that
case, we obtain for the domain wall resistance a strictly monotonic
increase with $\Del$ compatible to the findings of Brataas {\it et
  al}, in particular, there is no negative DWR. Another
puzzling fact appears, when we let the difference in momentum
scattering rates vanish, by setting $\gamma = 0$ in our result (let us
assume for a moment that $\gamma$ is an independent phenomenological
parameter), corresponding to $\tau_\UP = \tau_\DOWN$ in the result of
Brataas {\it et al} \cite{Brataas1999}.  We find
\begin{widetext}
  \begin{equation}
    f\left(\Del,\gamma=0,\xi=0,\kappa=0\right) = 
    \frac{\Del^3 \left(2+\Del^2\right)-P\left(4+\Del^2\right)}{2 \Del(1+P\Del)}
    = -\frac{3}{16} \Del^2 + O[\Del]^4\,,
  \end{equation}
\end{widetext}
which yields an overall leading order term proportional to
$\Del^2$. In contrast the result of Brataas {\it et al} yields a
constant contribution in that limit: $\delta\rho_{\rm W} = \frac 3 4
\frac {E_{\rm DW}} {\eF}$. Interestingly, exclusion of spin-flip
processes in our approach yields a result with the same asymptotics,
i.e. $\tilde f\left(\Del,\gamma=0,\xi=0,\kappa=0\right) = \frac{3}{4}
+ O[\Del]^2$. Unsurprisingly, the results still do not coincide
exactly, since we used the fully microscopic collision integral in our
approach and corrections to the electronic structure are already
present in this order.

\section{Conclusion and Outlook}
We have calculated the domain-wall resistance of a Bloch wall situated
in a wire of quasi one-dimensional geometry. The current flow is
perpendicular to the wall. We assumed that any spin-accumulation has
decayed at the contact ends, thus neglecting any finite size
effects. Going towards the half-metallic regime, our calculations show
the existence of a negative DWR in the presence of non-magnetic
scattering giving rise to spin-flip scattering. This possibility to
obtain a negative DWR is a band structure effect. Disagreement is
found when comparing our results to various previous works on DWR. We
believe these discrepancies seemingly arise, on the one hand, from
the neglection of spin-flip processes and, on the other hand ,from the
different approaches to include impurity scattering. While we use a
fully microscopic approach for scattering, reflected in the full form
of the collision integral (\ref{eq:CollisionIntegral}) with gradient
corrections and modification of the electronic structure properly
taken into account, other works introduced momentum scattering rates
phenomenologically.

To summarize, we have derived fully miscroscopic equation for the
spin transport in noncollinear magnetization textures. Our approach takes
impurity scattering and spin-flip scattering into account  on the
Hamiltonian level. This paves the way to treat more complex magnetic
textures and derive microscopic expression for the domain-wall-induced
resistance. 

All previous works dealing with DWR in the limit of wide walls obtain
results that depend in a similar way on the microscopic parameters,
that is, the DWR is $C \alpha^2$ where C is a dimensionless prefactor
and $\alpha=\frac{\hbar\vF}{\Delta w}$ is the spin mistracking
angle. Hence, every theory predicts this sort of dependency, but with
differnt proportionality factors, both in value and sign. This factor
$C$ however contains information about the scattering and DOS of the
two spin channels and can depend in a complex manner on these
properties. The most simple one is due to a model by Levy and Zhang
\cite{Levy1997} where $C$ simply depends on the ratio of resistivities
for spin up and down channels.

We did not take into account the possibility of magnetic moment
softening, i.e. the reduction of magnetic moment within the domain
wall. This effect is most prominent in very sharp domain walls where
canting of adjacent spin is large so that the noncollinear spin states
hybridize which in turn leads to a reduction in the absolute value of
the magnetic moment. As shown in \cite{Gorkom1999}, a reduction of the
magnetic moment can lead to a negative DWR.

Finally, effects due to geometric confinement have not been
considered, for example, surface scattering might become
important. Also, the magnetization profile can be more complicated and
might lead to eddy currents in the vicinity of the domain wall which
might be relevant for interpretation of experimental results on the
DWR in thin nanowires.

We thank Arne Brataas for discussions and acknowledge financial
support by the Deutsche Forschungsgemeinschaft through SFB 767 and SP
1285 and by the Landesstiftung Baden W\"urttemberg.

\appendix

\newcommand {\gx}[1] { {\hat g}^{(#1)} }
\newcommand {\ax}[1] { {\hat a}^{(#1)} }
\newcommand{\XP}  {\myTens {\hat \chi}}

\section{Appendix -- Details on deriving the hierarchy of equations}
We start out with the kinetic equation (\ref{eq:kineticEquationForg}) and (\ref{eq:CollisionIntegral}) by taking $(-e)\mv{v_x^n \  (\ref{eq:kineticEquationForg})}$ and by introducing
\begin{eqnarray*}
\hat g^{(n)} &=& (-e)\mv{v_x^n\ \hat g} = (-e) \int \frac {\upd^3 k} {(2\pi)^3} v_x^n \ \hat g \ , \\
\hat a^{(n)} &=& \mv{v_x^n\ \hat A} = \int \frac {\upd^3 k} {(2\pi)^3} v_x^n \ \hat A \ .
\end{eqnarray*}

Afterwards, employing an expansion of $\circ = e^{\frac i 2 ( \partialLeft{x} \partialRight{k_x} - \partialLeft{k_x} \partialRight{x})} \approx 1 + \frac i 2 ( \cdots) - \frac18 ( \cdots )^2$ and making use of relation (\ref{eq:TakingGAverages}), we obtain
\begin{widetext}
\begin{eqnarray}
\mv{v_x^n\ \frac i2 \left[ \hat f(x) \stackrel{\circ}{,} \hat g(x,k) \right]} &=& \frac i2 \left[\hat f, \gx{n}\right] + \frac {n} {2M} \frac12 \left\{ \px\hat f, \gx{n-1} \right\} - \frac{n(n-1)}{8M^2} \frac i2 \left[ \px^2 \hat f, \gx{n-2} \right] \ , \\
\mv{v_x^n\ \frac12 \left\{ \hat f(x) \stackrel{\circ}{,} \hat g(x,k) \right\}} &=& \frac12 \left\{\hat f, \gx{n}\right\} - \frac {n} {2M} \frac i2 \left[ \px\hat f, \gx{n-1} \right] - \frac{n(n-1)}{8M^2} \frac 12 \left\{ \px^2 \hat f, \gx{n-2} \right\} \ ,
\end{eqnarray}
where we are able to truncate the series since our treatment includes only terms up to order $q^2$ and subsequent terms would contribute only to higher orders. 

This yields the following equation
\begin{eqnarray}
\px\gx{n+1} - \Delta \frac i2 \left[\m\sig, \gx{n}\right] - \frac{\Delta n}{2M} \frac12\left\{\px\m\sig, \gx{n-1}\right\}
  + \frac{\Delta n(n-1)}{8M^2} \frac i2 \left[\px^2\m\sig,\gx{n-2}\right] - \frac n M (\px \delmu) \gx{n-1} \nonumber \\
 = \frac12 \left\{ \ax{n}, \XP \gx{0} \right\} - \frac12 \left\{ \XP \ax{0} , \gx{n} \right\} + \frac {n}{2M} \frac i2\left( \left[\ax{n-1}, \px\XP\gx{0} \right] + \left[ \px\XP\ax{0}, \gx{n-1} \right] \right) \nonumber\\
  -\frac{n(n-1)}{8M^2} \frac12 \left( \left\{ \ax{n-2},\px^2\XP\gx{0} \right\} - \left\{ \px^2\XP\ax{0}, \gx{n-2} \right\} \right) \ ,
\end{eqnarray}
where the action of $\XP$ is defined as
\begin{equation*}
\label{eq:XPOperation}
\XP \hat X = (\xii + \Tr\ximTens) \left(\hat X - \xi\kappa \ \vect x_\parallel \sig - \xi \ \vect x_\perp \sig \right)
 = \frac{\gjs}{2\pi\nu_0} \left(\hat X - \xi\kappa \ \vect x_\parallel \sig - \xi \ \vect x_\perp \sig \right)
\end{equation*}
\end{widetext}
and we write $\vect x_\parallel = \Tr\{\m\sig \hat X\} \m$ and $\vect x_\perp = \Tr\{\sig \hat X\} - \vect x_\parallel$, so that the total impurity self-energy, consisting of spin-isotropic and magnetic parts, equations (\ref{eq:SigmaImpurity}) and (\ref{eq:SigmaMagImpurity}), simply becomes
\begin{equation}
\check \Sigma = \check \Sigma_{\rm i} + \check \Sigma_{\rm mag} = \int \frac {\upd^3 k} {(2\pi)^3} \ \XP \check G \ .
\end{equation}

\newcommand {\nU}[1] {\nu_\UP^{(#1)}}
\newcommand {\nD}[1] {\nu_\DOWN^{(#1)}}
\newcommand {\nUp}[1] {{\nu_\UP'}^{(#1)}}
\newcommand {\nDp}[1] {{\nu_\DOWN'}^{(#1)}}
\newcommand {\nUpp}[1] {{\nu_\UP''}^{(#1)}}
\newcommand {\nDpp}[1] {{\nu_\DOWN''}^{(#1)}}
\newcommand {\nN}[1] {\nu_0^{(#1)}}
\newcommand {\nS}[1] {\nu_{\rm s}^{(#1)}}
\newcommand {\nNp}[1] {{\nu_0'}^{(#1)}}
\newcommand {\nSp}[1] {{\nu_{\rm s}'}^{(#1)}}
\newcommand {\nNpp}[1] {{\nu_0''}^{(#1)}}
\newcommand {\nSpp}[1] {{\nu_{\rm s}''}^{(#1)}}

\newcommand {\aP}[1]  {\alpha_\perp^{(#1)}}
\newcommand {\bN}[1]  {\beta_0^{(#1)}}
\newcommand {\bS}[1]  {\beta_{\rm s}^{(#1)}}
\newcommand {\bP}[1]  {\beta_\perp^{(#1)}}

We also need to know the spectral density $\ax{n}$ up to order $q^2$.
We find for even and odd moments, respectively
\begin{widetext}
\begin{eqnarray}
  \ax{2n} & = & 2\pi\left( \nN{n} \hat1 + \nS{n} \m\sig \right) +
  2\pi\left( \bN{n} \hat1 + \bS{n} \m\sig \right) (\px\m)^2 +
  2\pi\bP{n} (\px^2 \m)_\perp \sig \ , \nonumber \\ 
  \ax{2n+1} & = & 2\pi \aP{n} \m\!\times\!\px\m \sig \ ,
\end{eqnarray}
\end{widetext}
where $(\px^2 \vect m)_\perp$ denotes that we only take the component perpendicular to $\m$. The coefficients for a $d=3$ dimensional electron gas are given by
\begin{eqnarray*}
\aP{n} &=& \frac12 \left[ \frac2\Delta \nS{n+1} - \nNp{n+1} \right] \\
\bP{n} &=& -\frac1\Delta \aP{n} - \frac{2n(2n-1)}{8M^2}\nS{n-1} \\
\bN{n} &=& -\frac{2n-1}{4M} \aP{n-1} + \frac{2n-1}{M}\nN{n-1} \frac{\delmu}{q^2} \\
\bS{n} &=& -\frac{1}{2\Delta} \aP{n} - \frac{2n-1}{8M}\nS{n-1} \left[1 - \frac{8M}{q^2}\delmu \right] \ ,
\end{eqnarray*}
using the definitions
\begin{eqnarray*}
\nN{n}(\omega) &=& \frac12 \left[ \nU{n}(\omega) + \nD{n}(\omega) \right] \\
\nS{n}(\omega) &=& \frac12 \left[ \nU{n}(\omega) - \nD{n}(\omega) \right] \\
\nu_\UPDOWN^{(n)}(\omega) &=& \frac{\nu_\UPDOWN(\omega)}{1+2n} 
  \left[\frac{2}{M}\left(\omega \pm \frac\Delta2\right)\right]^n \ .
\end{eqnarray*}
$\nu_{\UPDOWN}$ specified without any argument implies that we
take its value at the Fermi-level, i.e. more specifically
$\nu_{\UPDOWN}^{(n)} \equiv \nu_\UPDOWN^{(n)}(\eF)$.

The chemical potential $\delmu$ is obtained from the condition that
\begin{widetext}
\begin{eqnarray}
\int_{-\infty}^{+\infty} \frac{\upd\omega}{2\pi} \ f_{\rm D}(\omega) \ \Tr\left\{\ax{0}(\omega) - \ax{0}(\omega) |_{q=0} \right\}
 = \int_{-\infty}^{+\infty} \upd\omega f_{\rm D}(\omega) \ 2\bN{0} q^2 = \nonumber\\
 \int_{-\infty}^{+\infty} \upd\omega f_{\rm D}(\omega) \ \frac{2}{M}\left(\frac{1}{4}\aP{-1}q^2 - \nN{-1} \delmu \right)
 = \frac{1}{2} \aP{0} q^2  + 2\nN{0} \delmu \stackrel{!}{=} 0 \ ,
\end{eqnarray}
since in the zero-temperature approximation
\begin{equation*}
\int_{-\infty}^{+\infty} \upd\omega \ f_{\rm D}(\omega) \nu_{\UPDOWN}^{(n)}(\omega) = \int^{\eF} \upd\omega \  \nu_{\UPDOWN}^{(n)}(\omega) = \frac{M}{2n+1} \nu_{\UPDOWN}^{(n+1)} \ .
\end{equation*}
\end{widetext}
Therefore, we immediately arrive at
\begin{equation}
\delmu = -\frac{1}{4\nN{0}} q^2 \aP{0}
 = \frac{q^2}{2M} \frac14 \left( 1 - \frac2\Delta \frac{M\nS{1}}{\nN{0}} \right) \ ,
\end{equation}
which yields equation (\ref{eq:chemicalPotential}), once we plug in all definitions and by noting that $\nN{0} = \nu_0$, $\nS{0} = P \nu_0$ and $\nS{1} = \frac{\nu_0}{3M}(2 P\eF+\Delta)$. Substituting $\delmu$ back into the expressions for $\beta$, we can now write
\begin{eqnarray*}
\bN{n} &=& -\frac{2n-1}{4M} \aP{n-1} - \frac{\nNp{n}}{4\nN{0}} \aP{0} \\
\bS{n} &=& -\frac{1}{2\Delta} \aP{n} - \frac1{4\Delta} \frac{\nS{1}}{\nN{0}} \nSp{n} \ .
\end{eqnarray*}

\begin{widetext}
Later, we will also need $$ \XP\ax{0} = \XP\ax{0}_0 + \XP\ax{0}_q \rightarrow \gjs (\hat1 + P(1-\xi\kappa) \m\sig)
  + \frac\gjs{\nu_0} \left( \bN{0} \hat1 + \bS{0} (1-\xi\kappa) \m\sig \right) q^2 + \frac{\gjs}{\nu_0}\bP{0} (1-\xi) (\px^2 \m)_\perp \sig  \ ,$$
\end{widetext}
where we write $\ax{0}_0$ to indicate that we take the zeroth order in $q$ only and accordingly, $\ax{0}_q$ contains all corrections to the density of states due to a magnetization gradient.

Next, we will change the $2\!\times\!2$ spin-matrix representation to the $4\!\times\!4$ matrix representation in the basis introduced above, $(+,-,\UP,\DOWN)$. This change of basis implies a local rotation of the basis in spin space to align the magnetization direction $\m$ along the new $z$-axis, which corresponds to the gauge transformation introduced above. The various vectors we will encounter in the following derivation transform in the following manner:
\begin{eqnarray*}
\m\sig &\rightarrow& \hat\sigma_z \\
\px\m\sig &\rightarrow& q \hat\sigma_y \\
\m\!\times\!\px\m &\rightarrow& -q \sigma_x \\
\px^2\m\sig = (\px^2\m)_\perp \sig + (\m\px^2\m)\m\sig &\rightarrow& (\px q)\hat\sigma_y - q^2 \hat\sigma_z \ ,
\end{eqnarray*}
and we note that $-\m\px^2\m = (\px\m)^2 = q^2$.
In the following, we specify substitution rules that perform this transformation:
\begin{equation}
\begin{array}{ll}
\frac i2 \left[ \hat\sigma_i , \hat n \right] & \rightarrow \myTens M_i \Vect n \\
\frac 12 \left\{ \hat\sigma_i , \hat n \right\} & \rightarrow \myTens K_i \Vect n
\end{array} \ ,
\end{equation}
where
\begin{widetext}
\begin{equation*}
\begin{array}{ccc}
\myTens M_x = \dfrac i2\left(\begin{array}{cccc}0&0&1&-1 \\ 0&0&-1&1 \\ 1&-1&0&0 \\ -1&1&0&0 \end{array}\right) &
\myTens M_y = \dfrac 12\left(\begin{array}{cccc}0&0&-1&1 \\ 0&0&-1&1 \\ 1&1&0&0 \\ -1&-1&0&0 \end{array}\right) &
\myTens M_z = \left(\begin{array}{cccc}-i&0&0&0 \\ 0&i&0&0 \\ 0&0&0&0 \\ 0&0&0&0 \end{array}\right) \\
\myTens K_x = \dfrac 12\left(\begin{array}{cccc}0&0&1&1 \\ 0&0&1&1 \\ 1&1&0&0 \\ 1&1&0&0 \end{array}\right) &
\myTens K_y = \dfrac i2\left(\begin{array}{cccc}0&0&1&1 \\ 0&0&-1&-1 \\ -1&1&0&0 \\ -1&1&0&0 \end{array}\right) &
\myTens K_z = \left(\begin{array}{cccc}0&0&0&0 \\ 0&0&0&0 \\ 0&0&1&0 \\ 0&0&0&-1 \end{array}\right)
\end{array}
\end{equation*}
\end{widetext}
constitute matrices in $(+,-,\UP,\DOWN)$ representation.
The action of $\XP$ turns into $$ \XP \hat n \rightarrow \frac{\gjs}{2\pi\nu_0} \myTens\chi \Vect n$$
with the matrix $$ \myTens\chi = \myTens1 + \xi\left(\begin{array}{cccc}-1&0&0&0 \\ 0&-1&0&0 \\ 0&0&-\frac\kappa2&\frac\kappa2 \\ 0&0&\frac\kappa2&-\frac\kappa2 \end{array}\right) \ . $$

As a consequence of the gauge transformation, the derivative transforms into
\begin{equation*}
\px \rightarrow \myTens\px \equiv \myTens 1 \px + q \myTens M_x \ .
\end{equation*}

\newcommand {\gz}[1] { {\Vect g}^{(#1)} }
Now with these rules at hand, the change of representation is straightforward and we obtain
\begin{widetext}
\begin{equation}
\myTens\px \gz{n+1} + \Gaj \gz{n} = \myTens\Xi_0^{(n)} \gz{0} + \myTens\Xi_q^{(n)}\gz{0} - \myTens\Gaj_q^{(n)} \gz{n} + \myTens\zeta_1^{(n)} \gz{n-1} +\myTens\zeta_2^{(n)} \gz{n-2}
\end{equation}
where zeroth order relaxation and precession terms are
\begin{equation}
\frac12 \left\{ \XP \ax{0}_0 , \gx{n} \right\} - \Delta \frac i2 \left[\m\sig, \gx{n}\right] \rightarrow
\left[ \gjs(\myTens1 + P(1-\xi\kappa)\myTens K_z) -\Delta \myTens M_z \right]\gz{n} \equiv \Gaj \gz{n}
\end{equation}
and magnetization gradient correction to relaxation rates yield
\begin{equation}
\frac12 \left\{ \XP \ax{0}_q , \gx{n} \right\} \rightarrow \frac{\gjs}{\nu_0}\left[q^2\bN{0}\myTens1 + q^2 \bS{0}(1-\xi\kappa)\myTens K_z + (\px q)\bP{0}(1-\xi) \myTens K_y\right]\gz{n} \equiv \myTens\Gaj_q^{(n)} \gz{n} \ .
\end{equation}
Corrections that depend on lower moments $\gz{n}$ in the hierarchy read explicitly
\begin{eqnarray}
\frac{\Delta n}{2M} \frac12\left\{\px\m\sig, \gx{n-1}\right\} + \frac n M (\px \delmu) \gx{n-1}
  + \frac {n}{2M} \frac i2 \left[ \px\XP\ax{0}, \gx{n-1} \right] \nonumber\\
 \rightarrow \left[q\frac{\Delta n}{2M} \myTens K_y + \frac n M (\px \delmu)\myTens1 + q\frac {n}{2M}\gjs\gamma \myTens M_y\right] \gz{n-1} \equiv \myTens\zeta_1^{(n)} \gz{n-1}
\end{eqnarray}
and
\begin{eqnarray}
-\frac{\Delta n(n-1)}{8M^2} \frac i2 \left[\px^2\m\sig,\gx{n-2}\right] + \frac{n(n-1)}{8M^2} \frac12 \left\{ \px^2\XP\ax{0}, \gx{n-2} \right\}  \nonumber\\
 \rightarrow \frac{n(n-1)}{8M^2} \left[\Delta \left(q^2\myTens M_z - (\px q)\myTens M_y\right) - \gjs\gamma \left(q^2\myTens K_z - (\px q)\myTens K_y\right) \right] \gz{n-2} \equiv \myTens\zeta_2^{(n)} \gz{n-2} \ .
\end{eqnarray}
\end{widetext}
Furthermore, we have source terms appearing in the equation, whereof the zeroth order term simply is
\begin{equation}
\label{eq:XiDefinition}
\frac12 \left\{ \ax{2n}_0, \XP \gx{0} \right\} \rightarrow \frac{\gjs}{\nu_0} \left[ \nN{n} \myTens1 + \nS{n}\myTens K_z \right] \myTens\chi \gz{0} \equiv \myTens\Xi_0^{(2n)} \gz{0}
\end{equation}
for even indices and $\myTens\Xi_0^{(2n+1)} = 0$ for odd indices.
The corresponding gradient corrections are, for even indices
\begin{widetext}
\begin{eqnarray}
\frac12 \left\{ \ax{2n}_q, \XP \gx{0} \right\} + \frac {2n}{2M} \frac i2\left[\ax{2n-1}, \px\XP\gx{0} \right]
    -\frac{2n(2n-1)}{8M^2} \frac12 \left\{ \ax{2n-2},\px^2\XP\gx{0} \right\} \nonumber\\
  \rightarrow \frac{\gjs}{\nu_0}\left[ q^2\bN{n}\myTens1 + q^2\bS{n}\myTens K_z + (\px q) \bP{n} \myTens K_y
  -q \frac {2n}{2M} \aP{n-1} \myTens M_x \myTens\px -\frac{2n(2n-1)}{8M^2} \left(\nN{n-1}\myTens1 + \nS{n-1}\myTens K_z\right) \myTens\px^2 \right] \myTens\chi \gz{0} \nonumber\\
  \equiv \myTens\Xi_q^{(2n)}\gz{0} \ .
\end{eqnarray}
and for odd indices
\begin{eqnarray}
\frac12 \left\{ \ax{2n+1}_q, \XP \gx{0} \right\} + \frac {2n+1}{2M} \frac i2\left[\ax{2n}, \px\XP\gx{0} \right]
  \rightarrow \frac{\gjs}{\nu_0}\left[-q\aP{n} \myTens K_x + \frac {2n+1}{2M} \nS{n} \myTens M_z\myTens\px \right] \myTens\chi \gz{0}
  \equiv \myTens\Xi_q^{(2n+1)}\gz{0} \ .
\end{eqnarray}
\end{widetext}

\newcommand{\PiI} {\myTens\Pi^{-1} }
\newcommand{\PiD} {\myTens{\Pi D} }
\newcommand{\Mx} {\myTens M_x}
\newcommand{\myP}[2] {\myTens \Pi_{#1}^{(#2)} }
\newcommand{\myX}[2] {\myTens \Xi_{#1}^{(#2)} }
\newcommand{\myZ}[2] {\myTens \zeta_{#1}^{(#2)} }
\newcommand{\myY}[1] { \myTens Y_\mathsmaller{#1} }

For our purpose, we only need the first 5 equations, since, as stated previously,
in our regime of investigation we need to know only up to $\gx{4}$. Explicitly, these equations read
\begin{eqnarray}
\label{eq:HierarchyZero}
\myTens\px \Vect j + (\myTens\Pi - \myX{0}{0})\Vect n &=& (\myX{q}{0} - \myP{q}{0}) \Vect n \\
\label{eq:HierarchyOne}
\myTens\px \Vect S + \myTens\Pi \Vect j &=& \myX{q}{1} \Vect n - \myP{q}{1} \Vect j + \myZ{1}{1} \Vect n \\
\label{eq:HierarchyTwo}
\myTens\px \Vect T + \myTens\Pi \Vect S - \myX{0}{2}\Vect n &=& \myX{q}{2}\Vect n - \myP{q}{2}\Vect S + \myZ{1}{2}\Vect j + \myZ{2}{2}\Vect n \nonumber\\ \\
\label{eq:HierarchyThree}
\myTens\px \Vect U + \myTens\Pi\Vect T &=& \myX{q}{3}\Vect n + \myZ{1}{3}\Vect S + O(q^3)\\
\label{eq:HierarchyFour}
\myTens\Pi \Vect U - \myX{0}{4}\Vect n &=& O(q) \ ,
\end{eqnarray}
where we defined $\Vect n = \gz{0}$, $\Vect j = \gz{1}$, $\Vect S = \gz{2}$, $\Vect T = \gz{3}$ and $\Vect U = \gz{4}$.
Here, we already dropped terms that would only contribute to higher orders than $q^2$. Note that $\Gamma = \myTens\Pi - \myX{0}{0}$.

Our aim is to obtain a differential equation of the form (\ref{eq:finalDGLforN}), 
\begin{equation}
\label{eq:DGLforN}
\left(\Gan - \myTens D \px^2\right) \Vect n = \myTens W(q) \Vect n \ ,
\end{equation}
where
\begin{multline}
\label{eq:WDefinition}
\myTens W(q) = \myY{q\px} q\px + \myY{\px q} \px q + \\ \myY{\px qq\px} \px q^2 \px + \myY{q\px q\px} q\px q\px + \myY{\px q\px q} \px q\px q .
\end{multline}
To achieve this, we unite the set of equations (\ref{eq:HierarchyZero})-(\ref{eq:HierarchyFour}) iteratively by eliminating every moment except $\Vect n$. We do this order by order in $q,\px$ and in the following, we give only terms that are relevant to our result.
The zeroth order is simply $\Vect S'_0 = \PiI\myX02\Vect n$ and $\Vect U_0 = \PiI\myX04 \Vect n$ while the odd moments $\Vect j, \Vect T$ vanish as $q,\px\rightarrow0$.
$\Vect S_0$, when plugged into equation (\ref{eq:HierarchyOne}) will yield a term that resembles a diffusion term, but will not yet be of the desired form as given by (\ref{eq:DiffusionMatrix}).
In order to tranform the diffusion term into the form which permits convenient solution of the differential equation, we will multiply equation (\ref{eq:DGLforN}) to the left with $\myTens\Pi\myTens\calA$ and then subtract it from the equation (\ref{eq:HierarchyTwo}) which provides $\Vect S$.
We thus obtain a new $\Vect S_0 = \left( \PiI\myX02 + \myTens\calA\Gan \right) \Vect n$ with the choice of $\myTens\calA$ determined by the condition that $\Vect S_0 \stackrel{!}{=} \PiD\Vect n$ which in turn translates into
\begin{equation}
\myTens\calA = \left(\PiD - \PiI \myX{0}{2}\right)\myTens \Gamma^{-1}
\end{equation}
and, since $\myTens\Gamma$ is singular in longitudinal subspace as it possesses one vanishing eigenvalue, $\myTens\Gamma^{-1}$ has to be considered as the pseudo inverse.

In order to obtain the form reminiscent of the spin-charge diffusion
equation (\ref{eq:SpinContinuityDiffusionEquations}), we rewrite
equation (\ref{eq:HierarchyOne})
\begin{equation}
\label{eq:HierarchyOneAdjusted}
\myTens\Pi\left(\Vect j + \myTens D\px\Vect n \right) = \PiD\px \Vect n - \myTens\px \Vect S + \myX{q}{1} \Vect n - \myP{q}{1} \Vect j + \myZ{1}{1} \Vect n \ , \\
\end{equation}
where now the right-hand-side of this equation is of order q, in
particular the difference $\myTens\Pi\myTens D\ \Vect n - \Vect S$ no
longer contributes to the zeroth order solution, as opposed to $\Vect
S$ itself.

Before proceeding with higher orders, we define for convenience
\begin{eqnarray*}
\myP{q}{n} &=& \myP{qq}{n} q^2 + \myP{\px q}{n} (\px q - \px q) \\
\myZ{1}{n} &=& \myZ{1,q}{n} q + \myZ{1,\px qq}{n} (\px q^2 - q^2 \px) \\
\myZ{2}{n} &=& \myZ{2,qq}{n} q^2 + \myZ{2,\px q}{n} (\px q - \px q) \\
\myX{q}{2n} &=& \myX{qq}{2n} q^2 + \myX{\px q}{2n} \px q + \myX{q\px}{2n} q\px \\
\myX{q}{2n+1} &=& \myX{q}{2n+1} q + \myX{\px}{2n+1} \px \ ,
\end{eqnarray*}
where we remind that $\px$ acts on everthing to its right so that
$\myTens \Xi_q$ actually are differential opperators.

\begin{widetext}
The first order is given by ($\Vect j = \left(\myTens\calJ_\mathsmaller{q} q + \myTens\calJ_\mathsmaller{\px} \px\right)\Vect n$, etc.)
\begin{eqnarray*}
\myTens\calJ_\mathsmaller{q} &=& -\PiI \left(\Mx\PiD - \myX{q}{1} - \myZ{1q}{1} \right) \\
\myTens\calJ_\mathsmaller{\px} &=& -\myTens D + \PiI\myX{\px}{1} \\
\myTens\calT_\mathsmaller{q} &=& -\PiI \left(\Mx\PiI\myX{0}{4} - \myX{q}{3} - \myZ{1q}{3} \PiD \right) \\
\myTens\calT_\mathsmaller{\px} &=& -\PiI \left(\PiI\myX{0}{4} - \myX{\px}{3} \right) \ , \\
\end{eqnarray*}
the second order terms are
\begin{eqnarray*}
\myY{\px q} &=& \PiI\Mx\PiD + \myX{\px q}{0} - \myP{\px q}{0} - \PiI\left(\myX{q}{1} + \myZ{1q}{1}\right) \\
\myY{q\px} &=& \Mx\myTens D + \myX{q\px}{0} + \myP{\px q}{0} - \Mx\PiI\myX{\px}{1} \\
\myTens\calS_\mathsmaller{\px q} &=& \PiI \left(-\myTens\calT_\mathsmaller{q} - \myP{\px q}{2}\PiD + \left(\myX{\px q}{2} + \myZ{2\px q}{2} \right)\right) - \myTens\calA Y_{\px q} \\
\myTens\calS_\mathsmaller{q \px} &=& \PiI \left(-\Mx \myTens\calT_\mathsmaller{\px} + \myP{\px q}{2} \PiD + \myZ{1q}{2} \myTens \calJ_\mathsmaller{\px} + \left(\myX{q\px}{2} - \myZ{2\px q}{2}\right) \right) - \myTens\calA Y_{q\px}\\
\end{eqnarray*}
and finally, the 3 remaining coefficients in equation (\ref{eq:WDefinition}),
\begin{eqnarray*}
\myY{\px qq\px} &=& \PiI \left(\Mx \myTens\calS_\mathsmaller{q \px} + \myP{qq}{1} \myTens \calJ_\mathsmaller{\px} + \myZ{1\px qq}{1} \right) \\
\myY{q\px q\px} &=& \Mx\PiI \left(\myTens\calS_\mathsmaller{q \px} + \myP{\px q}{1} \myTens \calJ_\mathsmaller{\px} \right) \\
\myY{\px q\px q} &=& \PiI \left(\Mx \myTens\calS_\mathsmaller{\px q} - \myP{\px q}{1} \myTens \calJ_\mathsmaller{q} \right) \ .
\end{eqnarray*}

To solve equation (\ref{eq:DGLforN}), we use the method elaborated upon previously in this article, so that by use of equations (\ref{eq:formalSolutionlongitudinal}) and (\ref{eq:DWRGeneralFormula}) we end up with
\begin{eqnarray}
\delta \rho_{\rm DW} = \left[\frac{\myTensS K}{\lambda} \left[ 2 (\myY{\px q})_{\rm lt} \myTens \calF (\myY{q\px})_{\rm tl} + 
      (\myY{\px q})_{\rm lt} \myTens \calF (\myY{\px q})_{\rm tl} + (2 \myY{\px qq\px} + \myY{q\px q\px} + \myY{\px q\px q} )_{\rm ll} \right] \Vect n_0 \right]_{\rm c} \ ,
\end{eqnarray}
where the outer bracket denotes that we take the charge component.

\end{widetext}

\end{document}